\documentstyle[aps,prl,preprint]{revtex} 
\tighten
\input epsf
\newcommand{\E}{{\rm e}}
\newcommand{\D}{{\rm d}}
\newcommand{\I}{{\rm i}}

\renewcommand{\Re}{{\rm Re}}
\renewcommand{\Im}{{\rm Im}}

\begin{document}
\title{Non-linear effects and dephasing in disordered electron systems}
\author{R. Raimondi $^{(1)}$, P. Schwab$^{(2)}$, 
 and C. Castellani$^{(2)}$}
\address{ 
$^{(1)}$Istituto di Fisica della Materia e Dipartimento di Fisica "E. Amaldi",
 Universit\`a di Roma Tre, Via della Vasca Navale 84, 00146 Roma, Italy\\ 
$^{(2)}$Istituto di Fisica della Materia e Dipartimento di Fisica,
 Universit\`a ``La Sapienza'', piazzale A. Moro 2, 00185  Roma - Italy\\}

\date{\today}
\maketitle
\begin{abstract}
The calculation of the dephasing time in electron systems is presented.
By means of the Keldysh formalism we discuss in a unifying way both weak
localization and interaction effects in disordered systems. This allows us to
show how dephasing arises both in the particle-particle channel
(weak localization) and in the 
particle-hole channel (interaction effect).
First we discuss dephasing by an external field.
Besides reviewing previous work on how an external oscillating field suppresses
the weak localization correction, we derive a new expression for the 
effect of a field on the interaction correction.
We find that the latter may be suppressed by a static electric field,
 in contrast
to weak localization. We then  consider dephasing due to inelastic scattering.
The ambiguities involved in the definition of the dephasing time are clarified
by directly comparing the diagrammatic approach with the path-integral approach.
We show that different dephasing times appear in the particle-particle
and particle-hole channels.
Finally we comment on recent experiments.
\end{abstract}

\section{Introduction}

Quantum corrections to the classical Drude formula for the electrical
conductivity 
give rise  to singular terms in low-dimensional systems\cite{altshuler85,lee85}.
There are  two types of such terms. The first is known as the
weak localization correction
(WL) and arises as a result of the quantum interference of electron waves
in disordered systems. The second is a consequence
of the enhancement of the  electron-electron
interaction in a disordered system and is usually referred to as
electron-electron interaction correction (EEI).
In two dimensions, in particular, the correction to the conductivity
is logarithmic for both the WL and EEI correction.
The argument of the logarithm contains the  ratio of two length scales.
The first is  the mean free path, $l$, which sets the microscopic scale,
beyond which the system behaves diffusively.
The second length scale differs in the two cases. For WL it is $L_{\phi}$,
the scale over which
inelastic scattering starts to destroy the interference effects.
For EEI it is usually given by the thermal length $L_T$.
In a diffusive system, all length scales correspond to
characteristic times: the elastic scattering time $\tau =l^2 /D$,
the dephasing time $\tau_{\phi}=L^2_{\phi}/D$, and
the thermal time $\tau_T=\hbar /k_B T =L^2_{T}/D$.
In general, dephasing occurs either as a result of the interaction of the
system with the external environment or as a consequence of the internal
electron interactions. For example, it is
known that an AC-field suppresses the WL. On the other hand,
while it is clear that inelastic scattering contributes to dephasing,
the exact way this happens is by far less obvious
as witnessed by the recent hot debate in the literature
\cite{golubev98,critic,altshuler98}.
Natural questions to ask are: How does the dephasing time $\tau_{\phi}$
that enters the WL correction and cuts off the 
logarithmic singularity depend on inelastic scattering and on the
external environment?
Does dephasing also affect the EEI correction?
The two issues need to be addressed together if one wants to
understand the experimental data in detail.
Although one can find in the literature discussions of both these issues,
there seems to exist no unified discussion of them.
It is one of the aims of the present paper to fill this gap.

It was first noted by Schmid\cite{schmid74} that the inelastic quasiparticle 
scattering time is enhanced in the presence of disorder. 
The dephasing time, which controls the WL correction was initially
assumed to be identical to the inelastic quasiparticle
scattering time in a disordered Fermi-liquid \cite{abrahams81}.
According to this analysis, the inverse dephasing time was assumed 
to be $\propto T\ln T$ in 2d, thus predicting a violation of the Fermi-liquid
behavior at low temperatures.
However,  Altshuler, Aronov, and Khmelnitsky\cite{aak82}
(hereafter referred to as
AAK), by means of a semi-classical path-integral approach to the WL
correction, were able to
calculate the dephasing time directly and predicted an inverse
dephasing time proportional to $T$, in contrast to Ref.\cite{abrahams81}.
Some time later Fukuyama and Abrahams\cite{fa83} (hereafter indicated as FA),
re-examined the problem in terms of standard
diagrams and calculated the ``mass'' term that develops
in the particle-particle propagator. They found the same result
as Ref.\cite{abrahams81}. The origin of the discrepancy of the AAK and FA
results lies in a genuine
ambiguity in the definition of the dephasing time itself.
A review of the attempts to clarify this issue made
at the time\cite{fukuyama84,eiler84,aronov84} may be found in \cite{abrahams85}.
The issue was further complicated from the fact that
a dephasing time also appears in the particle-hole channel\cite{castellani86}.
The mass term in the particle-hole propagator turned out
to coincide with that found by FA in the particle-particle propagator.

More recently inelastic scattering and dephasing have been addressed from
a general point
of view by Stern et al \cite{stern90} and in connection with disordered
mesoscopic systems by various groups\cite{blanter96}, in all cases
confirming the AAK result.
Furthermore, during the last couple of years,
various experimental groups have observed a saturation of the
dephasing time at low temperatures
\cite{lin87,pooke89,hiramoto89,mueller94,mohanty97}.
Usually this saturation has been related to heating effects or
to the presence of magnetic moments. In Ref.\cite{mohanty97},
however, these possibilities have been excluded
experimentally, and the observed saturation seems to be in contrast
to the available theories.
If the observed saturation of the dephasing time is due to an intrinsic
mechanism of disordered electron systems, this will have dramatic
consequences on the localization theory, such as preventing an insulating
ground state.
The above consideration led to
a fresh re-examination of the theory whose historical development
has been sketched above.
In particular, such an intrinsic mechanism has been claimed to be
discovered in Ref.\cite{golubev98}.
However in Refs.\cite{critic,altshuler98} this new theoretical result has been
questioned and at the same time in Ref.\cite{altshuler98} 
a conventional mechanism has been
suggested as a possible explanation of the experimentally observed
saturation of the dephasing time.
The authors of Ref.\cite{altshuler98} concluded that there is no need
to revise the existing theory of the dephasing time in WL.
While we completely agree with
Ref.\cite{altshuler98} on this latter point, we think that the interpretation
of the experiments is not yet settled and, especially, the effect of dephasing
on the EEI correction is not clear.

In this paper we
discuss dephasing in the particle-hole channel, and in parallel we
clarify the reason for the discrepancy in the AAK and FA calculations.
The difference between the results may be traced back to a different definition
of the dephasing time. In the FA case, the dephasing time is assumed to be the
mass of the particle-particle propagator as due to the self-energy corrections.
In this type of corrections, one does not  include diagrams 
which connect the upper
and the lower electron lines entering the particle-particle propagator.
In the AAK approach these diagrams, i.e. the vertex corrections, are 
taken into account.
The inclusion of these vertex diagrams
in the diagrammatic approach of FA leads to the AAK result\cite{ward}.
In the particle-hole channel, on the other hand, we find that vertex corrections
are negligible and the dephasing is entirely determined by the self-energy.
 
To see this we will adopt a real time-formulation based on the
Keldysh technique which allows us to discuss  WL and EEI 
corrections in a unified way.
Furthermore, in the real-time formulation non-linear
effects,  for example those due to an external field, can be incorporated
more easily. In order to address the issue of how dephasing arises
in the particle-particle and in the particle-hole channels, it is
instructive to start considering the situation of an external field.
This analysis, besides providing a relatively simple physical situation
to analyse has also a genuine experimental interest. For example,
there exists no satisfactory theoretical explanation for the
non-linear field effects observed in 
Refs.\cite{bergmann90,kravchenko96,yoon99,liu91,vitkalov88}. To this end
we will devote a substantial part of our analysis to the nonlinear effects
on WL and EEI corrections for which we shall find an unexpected difference
in their dependence on a static electric field.

The plan of the paper is the following. In the next section, by means of
the Keldysh formalism, we derive an expression for the additional current
in a disordered conductor due to both
the WL and EEI corrections. The formulae obtained are valid in the presence
of an arbitrary time-dependent external field. This result allows us, in
section III, to discuss the WL and EEI corrections in the presence
of both a DC and an AC electric field.
We show that, in the case of a DC field, the EEI correction is suppressed,
in contrast to what happens for the case of  WL correction. 
An AC field, on the
other hand,  suppresses both
the WL and EEI correction, but with a different dephasing time.
In section IV, we will review the FA calculation of the inelastic
scattering time in two dimensions and show how the AAK
result may be obtained from it. In section V the path integral approach
to the dephasing time of AAK 
is briefly outlined and connection is made with the diagrammatic analysis.
Special attention is paid to how  the cancellation of the infrared
singularity occurs. We extend this approach to a calculation of the dephasing time in
the particle-hole channel. Finally, section VI is devoted to 
some discussion of the experimental results.
In Appendix A we give technical details concerning the calculation of the dephasing time in
the particle-hole channel. 

\section{Quantum corrections to the conductivity}
\begin{figure}
\noindent
\hspace{0.5cm}{\epsfxsize=6cm\epsfysize=4cm\epsfbox{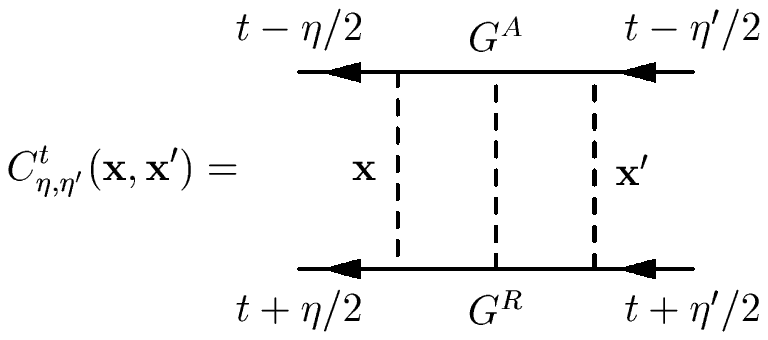}} \\
\hspace{0.5cm}{\epsfxsize=6cm\epsfysize=4cm\epsfbox{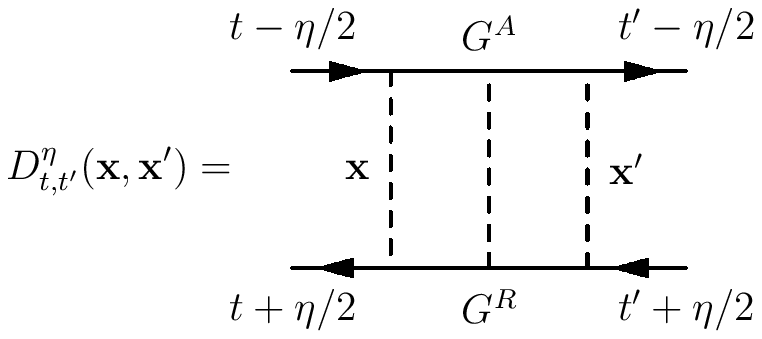}}

\vspace{1cm}
\caption{Graphical definition of the cooperon (particle-particle propagator)
and diffuson (particle-hole propagator). $G^A$ and $G^R$ are the advanced 
and retarded
Green's functions which in general depend on the external electromagnetic field.}
\label{fig2}
\end{figure}

In a weakly disordered metal, quantum corrections
lead at low temperature to deviations from the Drude-Boltzmann theory 
of transport.
The weak localization (WL) corrections are due to  electrons diffusing
along closed paths, where quantum interference causes an enhanced 
back-scattering probability.
The correction to the current in the sample is given by\cite{altshuler85}
\begin{equation}\label{WL}
\Delta {\bf j}_{\rm WL}(t) = -e^2 D \tau {4\over \pi}
\int_{\tau}^{\infty} \D \eta C^{t-\eta/2}_{\eta, -\eta}({\bf r},{\bf r})
{\bf E}(t-\eta)
,\end{equation}
where $D$ is the diffusion constant, $\tau$ the elastic scattering time
and $C^{t}_{\eta, -\eta}$
is the so-called cooperon or particle-particle propagator.
Its graphical definition is given in Fig.\ref{fig2}.
In the presence of
a vector potential the cooperon is given by the solution of the
differential equation  
\begin{eqnarray} 
&\left\{ 2 {\partial \over \partial \eta} +D
\left[ -\I \nabla + e {\bf A}({\bf r}, t-{\eta \over 2} ) +
                    e {\bf A}({\bf r}, t+{\eta \over 2}) \right]^2
\right\} C^t_{\eta \eta'}({\bf r }, {\bf r}' )
&\nonumber\\
&= {1\over \tau } \delta(\eta -\eta') \delta({\bf r } -{\bf r}' )
.&\end{eqnarray}
Here and below we use units where $\hbar = k_B =1$.
In addition to WL there are quantum corrections to the conductivity due to
the interplay of disorder and electron-electron interaction (EEI).
In the long-wavelength, low-frequency limit
the electron motion is diffusive, leading to an
enhancement of the effective interaction, 
$V({\bf q}, \omega) \to V({\bf q},\omega)/(-\I \omega + Dq^2 )^2$, and
 results
in either an enhanced or suppressed conductivity
in the case of the direct and exchange contribution
of the interaction, respectively.

In the following we derive these corrections to the conductivity in 
the presence of an external electromagnetic field.
Since one has to calculate a time-dependent, non-linear response function,
we work within the Keldysh formalism, using the notation of Ref.\cite{rammer86}.
This has the advantage that no analytic continuation is necessary at $T\ne 0$ in 
order to get the physical response functions. In particular the two-particle
 propagators 
like the cooperon
are directly defined in terms of a retarded and an advanced Green's function, and there is no
restriction on the energies of these Green's function in contrast to what is found within
the Matsubara formalism.

The starting point is the general equation for the current in the presence
of both interactions and an external field. For fixed disorder configuration,
the current in $d$-dimensions is found from
the Keldysh component of the electron Green's function,
\begin{equation}
\label{eq3}\Delta {\bf j}_{\rm EEI}(t)=
\I \int{\D^d k \over (2 \pi )^d }{e {\bf k} \over m }
        \Delta G^K( t{\bf k}; t{\bf k} )
.\end{equation}
In order to simplify the notation, we only write down the time dependence of the
Green's function in the following.
To first order in the interaction  the correction to the 
Green's function is given by
\begin{eqnarray} \label{eq4}
\Delta G^K(t,t)& = &G^R(t,t_1) \Sigma^R(t_1, t_2) G^K(t_2,t) \nonumber\\
         && +G^K(t,t_1) \Sigma^A(t_1, t_2) G^A(t_2,t) \nonumber\\
         && +G^R(t,t_1) \Sigma^K(t_1, t_2) G^A(t_2,t)
,\end{eqnarray}
where $G$ is the non-interacting Green's function in 
the presence of both disorder and an external field, 
$\Sigma$ is the self energy, and one has to integrate 
over $t_1$ and $t_2$. It turns out, that only the
first two terms contribute to the quantum correction to the conductivity,
so the term $G^R \Sigma^K G^A$ will be neglected from now on.
The first two terms in eq.(\ref{eq4}) correspond, upon impurity averaging,
to the diagrams of Fig.2 of the first of Refs.\cite{altshuler80}.
To lowest order in the interaction the retarded self-energy is given by
\begin{eqnarray}\label{eq5}
\Sigma^R(t_1, t_2) &=& {\I \over 2} \left( 
    G^K(t_1, t_2)V^R(t_1, t_2) \right.\nonumber\\
&&\left. + G^R(t_1, t_2)V^K(t_1, t_2)
\right)
,\end{eqnarray}
where $V^R$ and $V^K$ denote the retarded and Keldysh component of the
interaction.
Singular corrections due to the interplay of disorder and interaction
arise, after averaging over disorder,
from singular vertex corrections, i.e. diffusons,
which appear on the vertices that connect an advanced and a retarded Green's
function.
Corrections to 
$\Delta G^K$ where both vertices are singular are much larger than the contributions 
with only one singular vertex. As a consequence, these various, less singular,
 terms
are neglected. In particular, the third term in
eq.(\ref{eq4}), and the second term in eq.(\ref{eq5}) give rise to only one
singular vertex and are therefore neglected.

\begin{figure}
\noindent
\hspace{0.5cm}{\epsfxsize=7cm\epsfysize=5cm\epsfbox{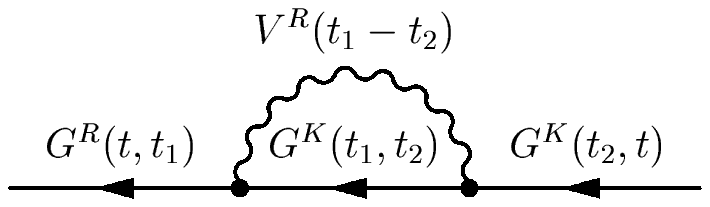}} 
(a) \\
\hspace{0.5cm}{\epsfxsize=7cm\epsfysize=5cm\epsfbox{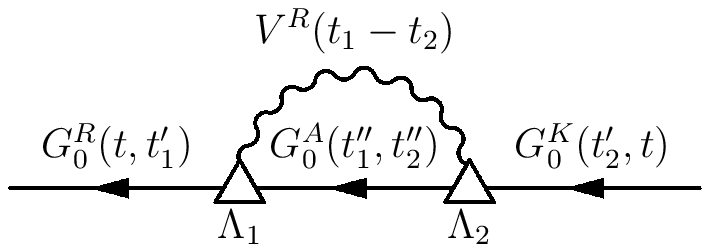}}
(b) \\
\hspace{0.5cm}{\epsfxsize=9cm\epsfysize=3.5cm\epsfbox{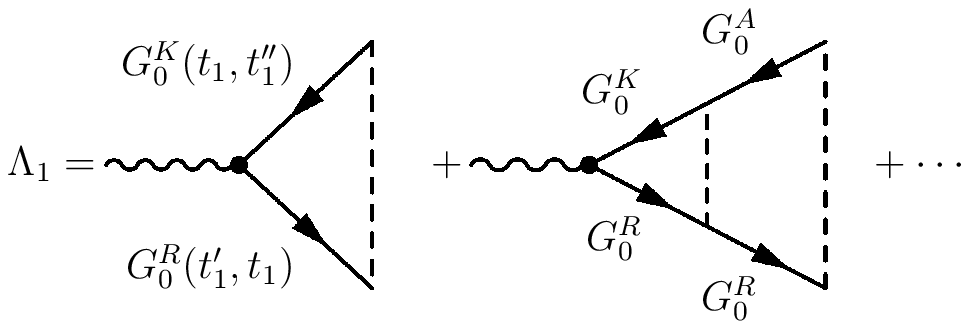}}
(c) 
\vspace{1cm}\\
\hspace{0.5cm}{\epsfxsize=9cm\epsfysize=3.5cm\epsfbox{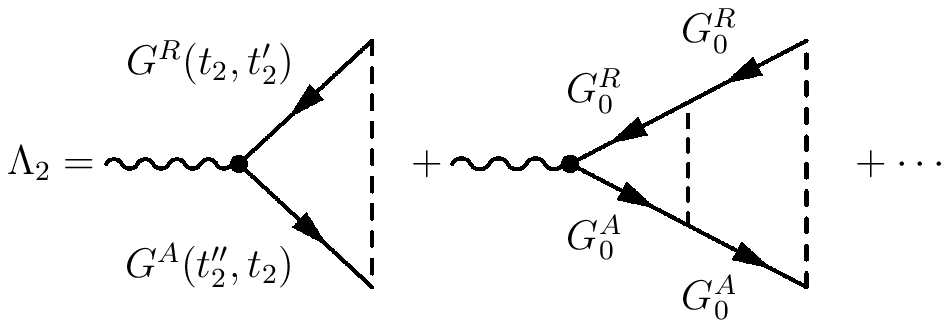}}
(d) \\

\vspace{1cm}
\caption{Diagrammatic structure of the correction to the Keldysh
component of the Green's function before (a) and after (b) averaging
over impurities. The vertices $\Lambda_1$ and $\Lambda_2$ are singular and
depend on the external electromagnetic field. Note the different
structure of $\Lambda_1$ and $\Lambda_2$: Whereas the bare interaction
vertex in $\Lambda_1$ is connected to a retarded and a Keldysh function,
the bare vertex in $\Lambda_2$ is directly connected to a retarded and an
advanced function.}
\label{fig1}
\end{figure}

The structure of the impurity averaged Keldysh Green's function is then
\begin{eqnarray}\label{eq6}
\Delta G^K(t,t) &=&
{\I \over 2}G^R_0(t, t_1') \Lambda_1( t_1', t_1''; t_1 ) G^A_0(t_1'', t_2'')
\nonumber\\
        && \times   V^R(t_1-t_2) \Lambda_2( t_2', t_2''; t_2)
           G^K_0(t_2', t) - c.c.
,\end{eqnarray}
where $G_0$ is the disorder average of $G$ and we have indicated by
$\Lambda_1$ and $\Lambda_2$ the vertices dressed by disorder.
Notice that $\Delta G^K$ is purely imaginary, and the current in eq.(\ref{eq3})
is real.
In order to make the structure of this equation clear, we represent 
$\Delta G^K(t,t)$ and the vertices $\Lambda_1$,
$\Lambda_2$ graphically in Fig.\ref{fig1}.
An external electromagnetic field affects all the Green's functions and all
the vertex functions in this expression.
However, it is known\cite{altshuler80} 
that the sum of the diagrams with two diffusons, 
i.e. the sum of the terms with an electromagnetic field insertion
in the Green's functions, 
cancel. Diagrams which contribute to the quantum corrections to the
conductivity must have at least three diffusons.
In our formalism these diagrams are generated by inserting 
the electromagnetic field
in the vertex corrections $\Lambda_1$ and $\Lambda_2$.
The singular part of the vertices is given by
\begin{eqnarray}
\Lambda_1(t_1', t_1''; t_1)& = &{1\over 2 \pi N_0 \tau }
       \langle G^R(t_1', t_1) G^K(t_1, t_1'') \rangle \\
\Lambda_2(t_2'',t_2'; t_2 ) & = &{1\over 2\pi N_0 \tau}
       \langle G^A(t_2'', t_2) G^R(t_2, t_2') \rangle ,
\end{eqnarray}
where the brackets denote the impurity average.
$N_0$ is the single-particle density of states. 
$\Lambda_2$ can be easily expressed in terms of the diffuson,
\begin{equation}\label{eq9}
\Lambda_2(t_2'', t_2';t_2) = \delta(t_2'-t_2'') D^{t_2'-t_2''}_{t_2,t_2'}
\end{equation}
with
\begin{eqnarray}\label{eq10}
& \left\{ {\partial \over \partial t} + D
\left[ -\I \nabla + e {\bf A}({\bf r}, t+ {\eta \over 2} ) -
 e {\bf A}( {\bf r}, t-{\eta \over 2}) \right]^2 \right\}
D^{\eta}_{t,t'} &\nonumber \\
&= {1\over \tau} \delta( t-t' ) \delta ({\bf r } - {\bf r}' )
.&\end{eqnarray}
The graphical definition of the diffuson is given in
Fig.\ref{fig2}.
In comparing the diffuson and the cooperon in Fig.\ref{fig2},
it is worth recalling that the time evolution of the diffuson
is associated with the center-of-mass time 
of the particle-hole pair, whereas the relative time controls the evolution 
of the particle-particle pair of the cooperon.
It is also useful to compare the different way an external vector potential
enters in the equation for the diffuson and the cooperon. In the case of
the diffuson, the vector potential is felt only if the particle and hole
are delayed with respect to one another, i.e. when $\eta \ne 0 $.
In this respect, 
processes, which are characterized by the simultaneous creation of a
particle-hole pair
in a given point in space, are not affected by
an external vector potential. This is the case 
for the density correlation function
which is relevant for screening the Coulomb interaction
and justifies the fact that the screened Coulomb interaction,
in eq.(15) below,
does not depend on the external vector potential.
Also the particle-hole pair, 
which appears in 
the vertex $\Lambda_2$, defined above, is composed of particles
created simultaneously.
$\Lambda_2$ therefore does not depend on the external vector potential.
By contrast the singular part of the vertex $\Lambda_1$ 
also involves the Keldysh component
of the disorder averaged Green's function, 
when it is expressed in terms of a diffusion propagator 
see Fig.\ref{fig2}c. 
The Keldysh function decays only slowly in time and introduces a time delay
between the creation of the particle and hole.
We approximate the Keldysh function by its equilibrium value
\begin{equation}\label{eq11}
G^K_0(\omega) = \tanh( \omega /2 T ) \left[ G^R_0(\omega ) - G^A_0(\omega) \right]
,\end{equation}
where $T$ is the electron temperature.
Notice that in strong electric fields where the electron temperature 
increases due to heating, the electron distribution function 
may differ from the Fermi
function, and eq.(\ref{eq11}) may break down.
Such effects are neglected in the present work.
Going back to the real-time formalism, we Fourier transform $G^K_0$.
The Fourier transform of
$\tanh(\omega /2 T)$ is $\delta(t) - \I T/\sinh(\pi T t )$, which
has to be convoluted with the advanced Green's function.
The delta function $\delta(t)$ 
can be neglected since it eventually leads to a term proportional to 
$\Theta (t_2-t_1)V^R(t_1, t_2)$ in the
self-energy $\Sigma(t_1,t_2)$ which is zero.
$\Lambda_1$ is then found as
\begin{equation}
\Lambda_1( t_1', t_1''; t_1 ) =
\int \D \eta D^\eta_{t_1' - \eta/2; t_1-\eta/2} {\I T \over \sinh (\pi T \eta )}
\delta(t_1'' -t_1'+\eta )
,\end{equation} 
and definitely depends on the external field due to the time delay $\eta$ 
in the particle-hole propagator.
Calculating the current requires also momentum integrations.
If the interaction $V$ transfers a momentum $\bf q$ with 
$ q  v_F\tau \ll 1$,
the $\bf k$-integration over the Green's functions gives
$$
\I \int {\D^d k \over (2 \pi )^d}  {e {\bf k } \over m }
G^R_0(t-t_1', {\bf k} ) G^A_0(t_1'-\eta-t_2',{\bf k}-{\bf q})
G^K_0(t_2'-t,{\bf k} )
$$
\begin{equation}
= 4 \pi N_0 \tau^2  \frac{T}{\sinh (\pi T \eta )} e D {\bf q }  \delta(t-t_1')
\delta(t_1'-\eta-t_2').
\end{equation}
The temperature dependent factor is due to $G^K(t_2'-t)$, with
$t_2'-t = -\eta$.  
Note that the vertex corrections $\Lambda_1$ and $\Lambda_2$ do not depend 
explicitly on ${\bf k}$ but only on the difference of momenta 
${\bf k} - ({\bf k } - {\bf q} )$.
Finally, we obtain the correction to the current due to 
the interplay of interaction 
and disorder as
\begin{eqnarray}
\Delta {\bf j}_{\rm EEI}(t) & =& - 
{ 4 N_0 \tau^2 e \over \pi }
\int \D t_1\D t_2 \D \eta  
\sum_{ \bf q}
D {\bf q}\left( { \pi T \over \sinh( \pi T \eta ) } \right)^2 \nonumber\\
&& V^R(t_1-t_2,{\bf q}) D^{\eta'=0}_{t_2, t-\eta}({\bf q})
D^{\eta}_{t-\eta/2, t_1-\eta/2}( \bf q)  \label{eq14}
.\end{eqnarray} 
Here for simplicity we assumed a homogeneous external field. In an inhomogeneous
field the diffuson is a function of two momenta,  
and the current is determined from
$D({\bf q}, {\bf q})$.
In the case of long range Coulomb interactions,  
the expression for the current simplifies.
In the relevant low momentum, low frequency region the
screened Coulomb interaction is given by
\begin{equation}\label{eq15}
V^R(\omega, {\bf q} ) ={1\over 2 N_0 }{-\I \omega + Dq^2 \over Dq^2 }
.\end{equation}
The frequency dependent numerator cancels when multiplied with the diffuson, 
so that the product
is frequency independent, and transforms into a delta function after Fourier
transformation. Thus the convolution of the interaction with 
the first of the two diffusons appearing in eq.(\ref{eq14}) gives
\begin{equation}
\int \D t_2 V^R(t_1-t_2, {\bf q} ) D^{\eta'=0}_{t_2, t-\eta}({\bf q})
= {1\over 2 N_0 \tau} {1 \over Dq^2} \delta(t_1-t+\eta )
,\end{equation}
and the expression for the current becomes
\begin{eqnarray} \label{EEI}
\Delta {\bf j }_{\rm EEI}(t) &=&
 -{2 \tau e \over \pi} \sum_{\bf q}\int_{\tau}^\infty
 \D \eta  { {\bf q } \over q^2}
\left( {\pi T \over \sinh( \pi T \eta ) } \right)^2
\nonumber\\
&& \times D^{\eta}_{t-\eta/2, t-3\eta/2 }({\bf q})
.\end{eqnarray}
This equation will be the starting point for our further investigations.
We cut off the time integration with $\tau$ since the diffusive
regime requires that all times are longer than $\tau$.
We see that as with $\Delta {\bf j}_{\rm WL}$,  
$\Delta {\bf j}_{\rm EEI}$ is also related to the propagation
of two particles. However, whereas WL is related to the propagation 
around a closed loop,
i.e. the probability of return $C({\bf r}, {\bf r})$, 
this is not the case for EEI,
due to the prefactor ${\bf q}/q^2$ in the momentum integration. 
Also the time dependencies are different. In the case of WL two particles start 
diffusing at $t-\eta$ at ${\bf r}_i$ and return at $t$ at the same point.
From the time dependence of the diffuson in eq.(\ref{EEI}), we conclude that,
in the case of EEI, the first particle starts at 
$t-2\eta$ at ${\bf r}_i$ and arrives at $t-\eta$ at ${\bf r}_f$.  
The second particle starts at $t-\eta$ at ${\bf r}_i$ and
arrives at $t$ at the final point ${\bf r}_f$.

Equation (\ref{EEI}) allows us to obtain the results known in the literature.
For instance, the interaction correction to the conductivity 
is found by expanding 
eq.(\ref{EEI}) to first order in the electric field.
By using eq.(\ref{eq10}) one finds the diffuson to linear order
in the electric field as
\begin{equation}
D^\eta_{t-\eta /2, t-3\eta /2}({ \bf q }) = {1\over \tau}
\E^{-Dq^2 \eta }\left( 1 + 2 D e {\bf q}\cdot { \bf E} \eta^2 +\cdots
\right)
\end{equation}
and one arrives at
\begin{equation} \label{eq19}
\Delta \sigma_{\rm EEI} =  - { 4e^2 D \over \pi } {1\over d }
\int_\tau^{\infty}  \D \eta
\left( {\pi T \eta \over \sinh (\pi T \eta ) }\right)^2
\left( {1\over  4 \pi D \eta }\right)^{d/2} 
,\end{equation}
where $d$ is the dimension.
The hyperbolic sine provides a cutoff for large times $\eta$.
By approximating  the temperature dependent factor with $1$ for $\eta T <1 $
and $0$ for $\eta T > 1 $, one finds immediately the temperature dependence
of the correction to the conductivity as $\Delta \sigma_{\rm EEI} \propto 1/\sqrt{T}$,
$\ln T$ and $\sqrt{T}$ in $d=1,2,3$. In one and three dimensions, the
correct prefactor is only found by integrating eq.(\ref{eq19}) with the result
\begin{equation} \label{eq22}
\Delta\sigma_{\rm EEI} \approx
\left\{ \begin{array}{ll}
- 2.5 \cdot {2 e^2 \over \pi} \sqrt{ D/ T }
& (d=1) \\
- {e^2 \over 2 \pi^2  } \ln( 1/T\tau ) & (d=2) \\
1.8  \cdot {e^2 \over 6 \pi^2 } \sqrt{ T/ D }
& (d=3)
,\end{array}
\right.
\end{equation}
where we subtracted temperature independent terms.
Here we only considered the contribution to $\Delta \sigma_{\rm EEI}$ from
the exchange diagram shown in Fig.\ref{fig1}. A correction to the conductivity
with the same temperature dependence as in eqs.(\ref{eq22})
is also found from the direct (Hartree) diagrams,
as shown in the literature\cite{altshuler85}. These will then change the total amplitude of
the quantum correction to the conductivity due to the interplay of disorder
and interaction. We will not consider this effect here and below.

For completeness, we recall also the results for the WL correction to
the conductivity in a weak static field.
The cooperon is given by
\begin{equation}
\label{cooperone}
C^t_{\eta, -\eta}({\bf r},{\bf r} ) = {1\over 2 \tau}
\left( {1\over 4 \pi D \eta } \right)^{d/2} \E^{-\eta/\tau_\phi }
.\end{equation}
Here a phenomenological dephasing time $\tau_\phi$ has been introduced in order
to cut off the infrared singularity.
The microscopic origin of such a dephasing time will be discussed later.
By inserting eq.(\ref{cooperone}) in eq.(\ref{WL}), one arrives at the well known expressions
\begin{equation}
\Delta\sigma_{\rm WL} = \left\{ \begin{array}{ll}
-{e^2 \over \pi }  \sqrt{ D \tau_\phi }
& (d=1) \\
-{e^2\over 2\pi^2 } \ln(\tau_\phi / \tau )  & (d=2) \\
{e^2 \over 2 \pi^2 }
\sqrt{1/ D \tau_\phi} & (d=3).
\end{array} \right.
\end{equation}

\section{Dephasing by a homogeneous electric field}
The quantum corrections to the conductivity may be suppressed due to
various mechanisms. We start by studying the effect of a homogeneous external
electric field.
The effect of an electromagnetic
field on WL has been described in the literature\cite{altshuler85,lee85}.
Here we consider in addition the effect of such a field on EEI.
We will first consider the non-linear response to a strong DC electric field, then
we will discuss the response to a DC field in the presence of a microwave field.

From eqs.(\ref{WL})
and (\ref{EEI}) both the linear and the non-linear response due to quantum 
interference can be determined. Note that heating effects are neglected in the
present discussion. In a static electric field, the vector potential
is given by ${\bf A}(t) = -{\bf E} t$.
The cooperon is then found as
\begin{eqnarray}
C^t_{\eta, -\eta }({\bf r}, {\bf r } ) &=&
{1\over 2 \tau} \sum_q \exp \left[ - D( {\bf q}- 2e {\bf E} t )^2 \eta \right]
\nonumber\\
&= &{1\over 2 \tau} \sum_q \exp \left[ - Dq^2 \eta \right]
,\end{eqnarray}
i.e. the electric field drops out. As a consequence, weak localization is not
affected by a static electric field\cite{altshuler85,lee85}.

There are, however, non-linear terms in the interaction correction to the 
current. The diffuson with the time arguments as needed for the calculation of the
current is given by
\begin{equation} \label{eq24}
D^{\eta}_{t-\eta/2, t-3/2\eta} ({\bf q})= 
{1\over \tau} \exp(-D({\bf q}- e{\bf E} \eta)^2\eta)
.\end{equation}
In contrast to weak localization, the interaction correction is not
related to a simple sum over the momentum entering  the two-particle propagator, since 
the factor ${\bf q }/q^2$ has to be taken into account. As a consequence, 
by shifting
the $q$ integration by $e {\bf E }\eta $ in eq.(\ref{eq24})
it does not simplify the calculation of the current, and especially 
there are non-linear electric field effects.
One can easily estimate beyond which electric field the non-linear 
effects become sizeable.
Equation (\ref{eq24}) can be expanded in the electric field under the condition
$ e E \eta < q$. On the other hand, 
as far as
the interaction correction to the current is concerned, only 
times $\eta < 1/T$ and momenta $q  > 1/L_T= (T/D)^{1/2}$ contribute effectively.
The electric field can thus be considered small when
$eE <T/ L_T$ which is equivalent to the condition
$D (eE)^2 \equiv T_0^3 < T^3$. Non-linear effects are expected to dominate 
for $e E  > T /L_T$, i.e.
when the
voltage drop over a thermal length exceeds the temperature.
For the sake of definiteness, let us consider eq.(\ref{EEI}) in two dimensions.
Performing the q-integration, the current reads
\begin{equation}
\label{cur2d}
\Delta {\bf j}_{\rm EEI} = - {\bf E }{ e^2 \over 2 \pi^2}
\int_{\tau}^{\infty}~ \frac{ d\eta}{\eta} 
\left(\frac{\pi T \eta }{\sinh \pi T \eta} \right)^2
\frac{\sinh (T_0^3 \eta^3/2 )}{T_0^3\eta^3/2}\exp(-T_0^3 \eta^3 /2 ).
\end{equation}
This does not lead to a simple linear dependence of the current on the electric field
and thus non-linear field effects are definitely present.
In the high temperature limit $T\gg T_0$, or weak electric field limit, it is 
the temperature dependent factor that  
cuts off the integral at large time $\eta$, whereas the $T_0$ dependent factor 
varies only weakly and is approximately one.
Including the first non-linear correction, the current is determined as
\begin{equation}\label{eq26}
\Delta {\bf j}_{\rm EEI} \approx -{e^2 \over 2\pi^2} {\bf E }
\left( \ln(1/T\tau) - 1.62 {D (e E)^2 \over (\pi T )^3 }
\right)
,\end{equation}
where $1.62$ is the approximate numerical factor from the $\eta$-integration.

In the limit of large electric field we approximate the $T_0$ dependent factor as
\begin{equation}
{\sinh( T_0^3 \eta^3 /2 ) \over T_0^3 \eta^3 /2 }
\exp( - T_0^3 \eta^3 /2 ) =
\left\{
\begin{array}{cc}
1 &  {\rm for}\, T_0 \eta < 1 \\
1/(T_0^3 \eta^3 ) & {\rm for }\, T_0 \eta > 1
\end{array}
\right.
\end{equation} 
and then 
we split the $\eta$-integration in eq.(\ref{EEI}) as
\begin{equation} \label{eq28}
\int_{\tau}^{1/T_0}\frac{d\eta}{\eta} +\int_{1/T_0}^{\infty}
\frac{\D \eta}{T_0^3\eta^4}  
\left( \frac{ \pi T \eta }{ \sinh (\pi T \eta) } \right)^2.
\end{equation}  
The second integral 
 gives only a small contribution to eq.(\ref{eq28}) provided that
$T_0 \tau \ll 1 $, and the value of the current at large fields
is found to be
\begin{equation} \label{eq25}
\Delta {\bf j}_{\rm EEI} = - {\bf E }{ e^2 \over 2 \pi^2} \ln(1/T_0 \tau)
.\end{equation}

\begin{figure}
\hspace{0.5cm}{\epsfxsize=10cm\epsfysize=5cm\epsfbox{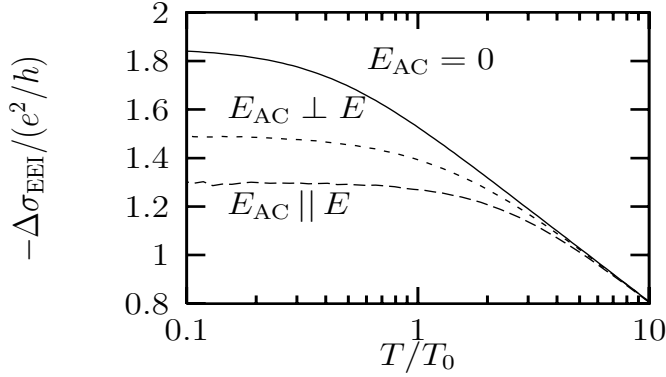}}

\vspace{1cm}
\caption{Temperature dependence of the conductivity
$\Delta \sigma_{\rm EEI} = \Delta j_{\rm EEI}/E$ due to the 
interplay of disorder and interaction in two dimensions.
The full line represents the current in presence of a static electric field
$E$. The dashed lines represents the conductivity, in presence of a
low frequency AC field of strength
$E_{\rm AC}=10E$ which superposes the static field $E$.
The AC field may be parallel or perpendicular to the DC electric field.
The temperature is in units of $T_0$, with $ D(eE)^2 = T_0^3$.}
\label{fig3a}
\end{figure}

Considering now the current as a function of temperature, one will observe
the $\ln T$ behavior for $T> T_0$ while the current will saturate to
eq.(\ref{eq25}) for $T<T_0$.
For illustration, we plot in Fig.\ref{fig3a} the non-linear conductivity 
as a function of 
temperature obtained by numerically integrating eq.(\ref{cur2d}).
The saturation of the conductivity at low
temperature is apparent.
In addition Fig.\ref{fig3a} also shows the conductivity 
in the presence of a slowly (in time) oscillating
field ${\bf E}_{\rm AC}$ which is superimposed to the current driving field $\bf E$.
This situation will be examined more closely below.
In Fig.\ref{fig3a} the conductivity is determined 
from the average of the current over one period
of the oscillating field divided by the static field. The 
strength of the oscillating field is $E_{\rm AC} = 10 E$.
At high temperature, the current is not affected by the oscillating field.
Since the oscillating field is much stronger than the DC field, 
the current saturates at much higher temperature than in absence of the AC field.
In more than one dimension one has to distinguish the two cases
${\bf E} \perp {\bf E}_{\rm AC} $ and ${\bf E} \, || \, {\bf E}_{\rm AC}$.
As it is apparent from the figure,
 a parallel AC field saturates the current at higher
temperature. 
If the frequency of the AC field is not too large ($\omega T < 1$),
the effect can be understood from eq.(\ref{eq25}) 
by considering the current in a strong electric
field ${\bf E}_{\rm total} = {\bf E}+ {\bf E}_{\rm AC}(t)$ and expanding
in the static field
\begin{equation}
\Delta {\bf j}_{\rm EEI} = { e^2 \over h}
\left[ {{\bf E}_{\rm AC}+{ \bf E} \over 3 \pi} \ln( e^2 D E_{\rm AC}^2 \tau^3 )
      +{2  {\bf E}_{\rm AC} \over 3 \pi} 
       { {\bf E}_{\rm AC} \cdot {\bf E} \over E_{\rm AC}^2 }
\right]
.\end{equation}
Upon time averaging the contribution, in the first term, 
not proportional to the DC field, vanishes.
The difference between parallel and perpendicular AC and DC field stems from the last term
and leads to $\Delta \sigma_{||} - \Delta \sigma_{\perp} = (2/3\pi )(e^2/h)$
in agreement with the numerical result in Fig.\ref{fig3a} 
in the low temperature region. 

\begin{figure}
\hspace{0.5cm}{\epsfxsize=10cm\epsfysize=5cm\epsfbox{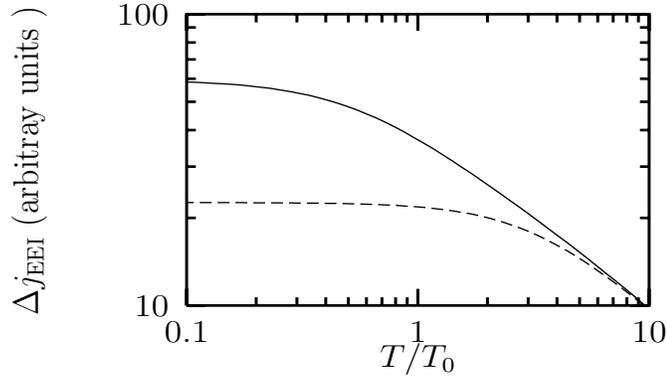}}

\vspace{1cm}
\caption{Temperature dependence of the additional current due to the
interplay of disorder and interaction ($d=1$).
The full line represents the current in presence of a static electric field
$E$. The dashed line represents the current, in presence of a 
low frequency AC field of strength
$E_{\rm AC}=10E$ which superposes the DC field $E$.
The AC field is assumed to be parallel to the DC field.
We use arbitrary units for the current,
and the temperature is in units of $T_0$, with $ D(eE)^2 = T_0^3$.
At high temperature, the $1/\sqrt{T}$ behavior which is characteristic 
in one dimension is seen. In this limit,
the current is linear in the electric field. Below
$\pi T \sim T_0 $, the current
saturates, due to the non-linear effects. In presence of the AC field the
current saturates at higher temperature.}
\label{fig3b}
\end{figure}

We find analogous results in one and three dimensions.
Figure \ref{fig3b} depicts the current as a function of temperature
obtained by numerically integrating eq.(\ref{EEI}) in one dimension. 
Again we studied the current in the presence of a static field only, and the
time averaged current in the presence of both a static and an oscillating field 
with $E_{\rm AC} = 10 E$. 
As in two dimensions, one
observes a saturation of the current in the low temperature region. 
The temperature scale, where the saturation occurs is roughly the same 
in one and two dimensions.

We consider now specifically the response in the presence of 
time-dependent electric fields.
The diffuson is then given by
\begin{eqnarray}
& D^{\eta}_{t-\eta/2,t-3\eta/2}({\bf q }) ={1\over \tau}
&\nonumber\\
&\times \exp
\left\{ -\int_{-\eta }^0 \D t_1 D \left(
{\bf q} + e{\bf  A}( t+t_1 ) - e {\bf A}(t+t_1-\eta ) \right)^2 
\right\}
.&\end{eqnarray}
We start with fields which are only slowly time dependent and discuss
high frequency fields later.
As we mentioned before, only small time delays, i.e. $\eta < 1/T$,
 are relevant for the calculation
of the interaction correction to the current. Therefore, when the vector potential
varies
only slowly on this time scale, it may be expanded in the equation above,
\begin{equation} \label{eq27}
{\bf  A}( t+t_1 ) - {\bf A}(t+t_1 -\eta )
\approx - \eta {\bf E} (t) 
, \end{equation} 
and one finds that the current instantly follows the electric field in this case.
The conductivity in the presence of a slowly oscillating field in Figs.\ref{fig3a} and
\ref{fig3b} was determined using this approximation.

In the case of a fast time dependence of the vector potential,
 the approximation (\ref{eq27}) 
is not sufficient and the full time dependence has to be taken into account. 
For ${\bf A} = {\bf E}_{\rm AC} \cos (\omega t )/\omega $ the relevant 
combination is
\begin{equation} \label{eq33} 
{\bf A} (t_1 +\eta/2) - {\bf A}( t_1 -\eta/2 ) =
- 2{\bf E }_{\rm AC}  \sin( \omega t_1 ) \sin(\omega \eta/2 )/\omega 
.\end{equation} 
In the high frequency limit the sine-functions are of order one, so that 
compared to (\ref{eq27})
the vector potential is reduced by a factor $\omega\eta$.
The linear response to a high frequency electric field, for instance,
becomes temperature independent
for $\omega \gg T$, since the main contribution to
$\Delta\sigma_{\rm EEI}$, eq.(\ref{EEI}), comes from $\eta < 1/\omega \ll 1/T$ and
processes with $\omega \eta >1$ are practically cut off.
The conductivity is then found as 
$\Delta \sigma_{\rm EEI} \propto \omega^{1/2}$, $\ln(\omega \tau)$, $\omega^{-1/2}$ in
three, two and one dimension, as it is discussed in the literature \cite{altshuler85}.

\begin{figure}
\hspace{0.5cm}{\epsfxsize=10cm\epsfysize=5cm\epsfbox{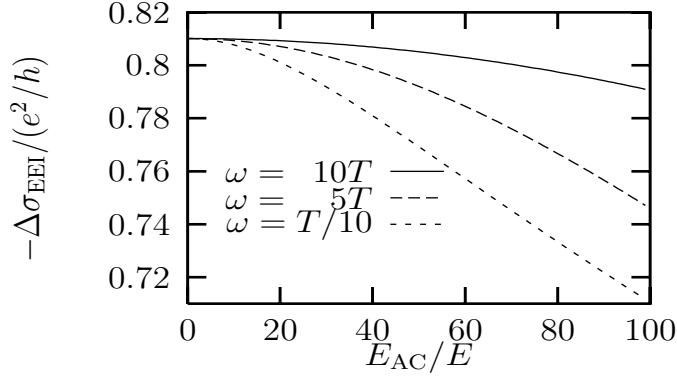}}

\vspace{1cm}
\caption{Conductivity in two dimensions in presence of both a DC field $E$
and a high frequency field $E_{\rm AC}$ of frequency $\omega$.
The temperature which is kept fixed is $T=10T_0$, where 
$T_0^3 = D(eE)^2 $. In the absence of the AC field one is in the
linear response regime, according to Fig.\ref{fig3b}. The AC field supresses
the quantum correction to the conductivity, but the effects becomes weaker
when $\omega > T$.}
\label{fig3c}
\end{figure}

We now concentrate on the effect of a high frequency radiation on the response  
to a static field.
In Fig.\ref{fig3c} we plot the static conductivity obtained by integrating
 eq.(\ref{EEI})
in the presence of both a static and a high frequency field in two dimensions. The static
and AC field are parallel.
The temperature is $T=10 T_0 $ with $T_0^3 = D (e E)^2$.
We find that the AC field reduces the quantum correction to the conductivity. 
The effect is strongest
for low frequencies, $\omega < T$, whereas in the limit of high frequencies the effect 
becomes weaker. 
In the relevant case of weak field and low frequencies, 
the correction to the static conductivity can be determined from (\ref{eq26})
with the result that
$\sigma_{\rm EEI}(E_{\rm AC})- \sigma_{\rm EEI}(0)
= 0.078 \cdot (e^2/2\pi^2) D(e E_{\rm AC})^2/T^3 $ if the AC and 
DC field are parallel and
one third of this for perpendicular AC and DC fields.
Concerning the high frequency limit we argued after
(\ref{eq33}) that the strength of the electric field in a high 
frequency is reduced by  a factor $\eta\omega$ 
which corresponds to $\omega/T$ if $\eta\sim 1/T$.
We have verified this estimate numerically and analytically and found  
that for week field the correction to the conductivity changes into
$ \sigma_{\rm EEI}(E_{\rm AC})- \sigma_{\rm EEI}(0)
\propto  D(e E_{\rm AC})^2/(\omega^2 T) $ in the high frequency limit.

To conclude this Section,
for the sake of completeness and comparison, 
we recall the results for the weak localization correction in a microwave
field\cite{altshuler85}. 
Notice that in the cooperon we need the sum of the vector potential at times
$t \pm \eta/2$,  
\begin{equation} 
{\bf A}_{\rm AC} ( t_1 +\eta/2) + {\bf A }_{\rm AC}(t_1-\eta/2 ) =
2 {\bf A}_{\rm AC} \cos(\omega t)
\cos(\omega \eta/2 )
,\end{equation} instead of the difference that enters the diffuson.
We average the cooperon over one period of the AC field and sum over momentum.
After some algebra one arrives at
\begin{eqnarray}
\Delta \sigma_{\rm WL} &= &  
-{2  e^2 \over \pi }{ D \over ( 4 \pi D)^{d/2} }
\int_\tau^\infty {\D\eta \over \eta^{d/2} }
\E^{-\eta/ \tau_\phi } 
\nonumber\\
&& \label{eq35}\times \exp\left[ {4 T_0^3 \over \omega^3 } B\left( {\omega\eta\over 2} \right) \right] 
   I_0\left[ {4T_0^3\over \omega^3 } B\left( {\omega \eta \over 2} \right) \right] \\
B(x)&=& x\left( 1+{\sin(2x)\over 2x }- 2 {\sin^2x \over x^2 } \right)
,\end{eqnarray}
where $T_0^3 = D(eE_{\rm AC})^2$ and $I_0(x)$ is the imaginary argument Bessel function.

The external microwave field does not affect the cooperon
for small $\eta$, but does so for large $\eta$.  
The estimate of the typical time scale $ \eta = \tau_{\rm AC}$, 
at which the external field modifies the $\eta$-integration in the equation above,
 is found 
from the condition that the argument of the exponential in the second line of
eq.(\ref{eq35}) is of order one, and one gets
\begin{equation}
1/\tau_{\rm AC} = \left\{ \begin{array}{ll}
2 D(e E_{\rm AC} )^2/ \omega^2   & {\rm for }\,\, D (e E_{\rm AC} )^2  \ll \omega^3  \\
\left[ D (e E_{\rm AC} )^2 \omega^2/180\right]^{1/5} 
& {\rm for }\,\,  D( e E_{\rm AC})^2 \gg \omega^3  
.\end{array}
\right.
\end{equation} 
Strong non-linear effects of the microwave field associated with WL will be seen 
when $\tau_{\rm AC}$ is smaller than the dephasing time $ \tau_{\phi }$.
For a fixed strength of the microwave $E_{\rm AC}$, the
time scale $\tau_{\rm AC}$ becomes large when  
$\omega \to 0$, $\tau_{\rm AC} \propto \omega^{-2/5}$, which
agrees with the observation that a static field does not
lead to nonlinear effects in WL.

\section{The inelastic scattering rate}
In this section we calculate the ``mass'' in the cooperon and
diffuson propagator.
This section does not contain  new material, since we will
reproduce the results of Refs.\cite{fa83,fukuyama84,castellani86}. 
However we think
that this section  
contains useful information in order to understand the different
approaches to the calculation of the dephasing time. 

In the presence of time reversal invariance,
the particle-particle propagator may be obtained from the particle-hole
propagator by reversing the direction of one of the electron lines.
As a consequence,
the mass term that develops in the particle-particle propagator
also enters the particle-hole propagator. A mass term in the particle-hole
channel could seem to imply, at first sight,
a violation of the particle conservation. However, this is not the case
because such a mass term eventually disappears in the physical response
function, as it has been shown by the detailed diagrammatic analysis of
Castellani et al.\cite{castellani86}.
We now review briefly the derivation of how such a mass term arises.
For our convenience, we adopt the notation of Ref.\cite{castellani86},
and restrict the discussion to the two-dimensional case.
The two-particle propagator is denoted by $L( {\bf Q}, \Omega )$
and may represent a cooperon or diffuson.

According to the standard diagrammatic technique,
a mass term for a propagator  may be obtained by considering
the appropriate self-energy. The self-energy diagrams for the two-particle
propagator are those labeled (a)-(d) in Fig.\ref{fig4}.
By denoting with $\Sigma({\bf Q} , \Omega_m ) $ the contribution
of these diagrams, one can use the Dyson equation to re-sum
their infinite series and obtain the two-particle propagator
$L({\bf Q},\Omega)$ as,
\begin{equation}
L({\bf Q}, \Omega_m ) = {1 \over |\Omega_m | + DQ^2 - 
\Sigma({\bf Q} , \Omega_m ) }.
\end{equation}
Successively, by expanding the
self-energy in powers of $\Omega$ and $Q^2$, one
may  identify the quantum corrections to the diffusion constant,
the density of states, and the frequency. This has been done explicitly
in Ref.\cite{castellani84}. 
The mass term, which will be called $1/\tau_{\rm inel}$, is found from
the self energy at zero momentum and frequency.
The only part of the graphs in Fig.\ref{fig4}, which is not manifestly
proportional to $Q$ or $\Omega$, is from graph (d) and is given by
\begin{equation}
{1\over{\tau_{\rm inel}}}=-{2}T
\sum_{\epsilon_n < \omega_{\nu} < \epsilon_n +\Omega_m}
\sum_{{\bf q}} V ({\bf q}, \omega_{\nu}) L ({\bf q}, \omega_{\nu}+\Omega_m ).
\label{eq36}
\end{equation}
Here, $\epsilon_n =\pi T (2n+1)$, $\omega_{\nu}=2\nu\pi T$, and 
$\Omega_m=2m\pi T$ are Matsubara frequencies with $\epsilon_n <0$ and
$\epsilon_n+ \Omega_m >0$. $\Omega_m$ is the external
frequency, which after analytical continuation ${\rm i}\Omega_m \rightarrow
\Omega +\I 0^+$, is sent to zero. In this limit, the number of addenda in the
Matsubara sum of eq.(\ref{eq36}) vanishes. However, the functions in the sum
contain a branch cut after the analytic continuation and eq.(\ref{eq36})
yields a finite result in the $\Omega \rightarrow 0$ limit.

After the analytic continuation with
${\I}\Omega_m\rightarrow \Omega +{\I}0^+$ and ${\I}\epsilon_n\rightarrow
\epsilon +{\I}0^-$, by taking the limit $\Omega, \epsilon \rightarrow 0$,
(i.e., particles at the Fermi surface),
and observing that the sum of the Fermi and Bose function 
$f(\omega )+b (\omega )=[\coth(\omega/2T)-\tanh(\omega/2T)]/2 =1/\sinh(\omega/T)$ is an 
odd function of $\omega$, one gets
\begin{equation}
{1\over \tau_{\rm inel}}=-{{2}\over{\pi}}\int_{-\infty}^{\infty}{\rm d}\omega
{1\over{\sinh{(\omega/T)}}}
\sum_{q} {\rm Re}L({\bf q},-{\rm i}\omega ) {\rm Im}V({\bf q},-{\I}\omega )
\label{6}
.\end{equation}
Considering the screened Coulomb interaction as  in eq.(\ref{eq15}), 
$\Im V({\bf q}, -\I \omega ) = -\omega/( 2 N_0 D q^2 )$,
the integral above is well behaved in the ultraviolet limit both
in frequency and momentum. Furthermore, due to the presence of the hyperbolic sine,
the main contribution to the inelastic scattering time
comes from  frequencies  $\omega < T$, physically corresponding to the fact
that there are no available excitations with larger energy. 
As usual, at a given temperature, only real excitations ($\omega < T$)
generate lifetimes, while virtual excitations ($\omega > T$) are immaterial.
On the other hand, the integral (\ref{6}) is infrared divergent
in one and two dimensions.
To cure this divergence we
insert the inelastic time self-consistently in the propagator
$L({\bf q}, \omega )$, and also calculate the screened Coulomb 
interaction more accurately.
In two dimensions one finds
\begin{equation}\label{eq40}
V({\bf q }, -\I \omega ) = {1\over 2 N_0} {\kappa \over q}
{Dq^2 -\I \omega \over Dq \kappa - \I \omega }
,\end{equation}
with $\kappa = 4 \pi e^2 N_0$, the inverse screening length.
Since the screening length is usually much shorter than the mean free path, 
eq.(\ref{eq40}) and eq.(\ref{eq15}) coincide in almost
 all the relevant parameter range,
but not in the extreme low momentum region, where there is no singularity in
$\Im V({\bf q}, \omega )$ according to eq.(\ref{eq40}).
By exploiting the fact that the main contribution 
to the frequency integral in (\ref{6}) comes from the
region
$| \omega | < T$, we may write in two dimensions
$$
{1\over \tau_{\rm inel}} = 
{ T \over 2 \pi^2 N_0 } \int_0^{\infty} \D q \int_{-T}^T \D \omega
{ D \kappa^2 q \over 
 \omega^2 + (D \kappa q )^2 } \nonumber
$$
\begin{equation}
\times { Dq^2 +1/\tau_{\rm inel }  \over
[\omega^2+ ( Dq^2 + 1/\tau_{\rm inel} )^2 ]}
.\end{equation}
The frequency dependence of the integrand is 
controlled by the two Lorentzian factors which have typical scales
$1/\tau_{\rm inel}$ and $D\kappa q $. Depending whether  
$D\kappa q > 1/\tau_{\rm inel}$ or $D \kappa q < 1/\tau_{\rm inel}$,
one can safely neglect the frequency dependence
of the first or second Lorentzian.
Performing then the $\omega$ integration,
we find 
$$
{1 \over \tau_{\rm inel} }
 = {T \over \pi^2 N_0 }  \int_{q> q_0} \D q 
   {1\over D  q} \arctan [ T/ ( Dq^2 +1/\tau_{\rm inel} ) ]
$$
\begin{equation}
+{T \over \pi^2 N_0 }  \int_{0}^{q_0} \D q \kappa \tau_{\rm inel}
\arctan (T/D\kappa q )
\end{equation}
where $q_0$ is given by $D q_0 \kappa = 1/\tau_{\rm inel }$.
For $1/\tau_{\rm inel} < T$, the dominant contribution to the inelastic scattering
rate comes from the momentum region $ q_0 < q < (T/D)^{1/2}$,
and we find
\begin{equation}\label{eq43}
{1 \over  \tau_{\rm inel }} =
{ T \over 4 \pi D N_0 } \ln( D \kappa^2 T \tau_{\rm inel}^2 )
\approx {T \over 4 \pi D N_0 } \log[ D \kappa^2 ( 4 \pi D N_0)^2 /T ]
\end{equation}
This is the FA result for the inelastic scattering rate we mentioned in the 
introduction. The AAK result for the dephasing time in
the weak localization can be obtained within the same calculation, but with a
different cut-off procedure of the low $q$ singularity.
One may argue that the Coulomb interaction cannot contribute to
dephasing on distances which are larger than the phase coherence length
itself, and therefore use $D q^2 = 1/\tau_\phi $ as a low momentum cut-off
in order
to determine the phase coherence length.
One then finds
\begin{equation}
\label{eq41}
{ 1\over \tau_\phi } = {T \over 4 \pi D N_0 } \ln( T\tau_\phi )
 \approx {T \over 4 \pi D N_0 } \ln(4 \pi D N_0 )
.\end{equation}
There remains the task to justify on a more formal level, the origin of the above
cut-off procedure.
Fukuyama suggested\cite{fukuyama84} to calculate $\Sigma( {\bf Q},\Omega )$ at 
finite $\bf Q$, so that the singularities in the propagator $L({\bf q}, \omega)$ 
and the interaction $V({\bf q}, \omega)$ in the calculation of the scattering rate
appear at different momenta, and their product is less divergent.
However this procedure does not cure the $1/q^2$ divergence of the interaction 
and thus cannot justify the difference between the FA and AAK results.
In the next section we will demonstrate how the low momentum cut-off
arises in case of weak localization within the path-integral
 approach of AAK\cite{aak82}.
We will also see that in the case of the interaction effect, no
such cancellation occurs, and the dephasing rates are different in the two
cases.
 
\section{Dephasing due to Coulomb interaction}
\begin{figure}
\hspace{0.5cm}{\epsfxsize=3cm\epsfysize=3cm\epsfbox{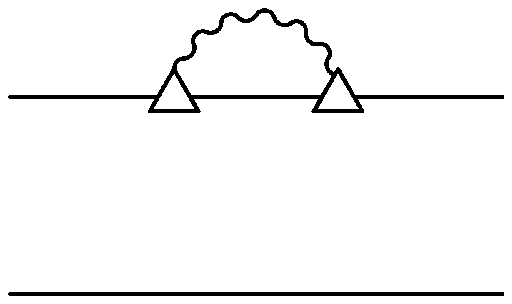}} (a)
\hspace{0.5cm}{\epsfxsize=3cm\epsfysize=3cm\epsfbox{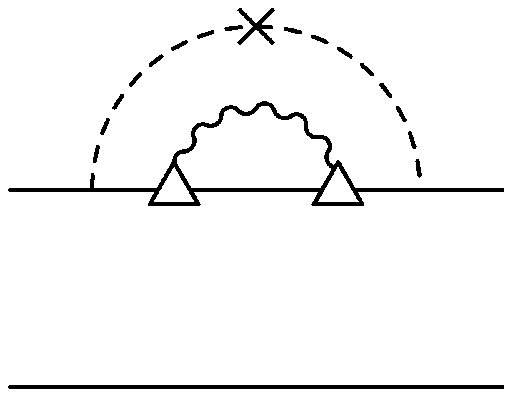}} (b)\\
\hspace{0.5cm}{\epsfxsize=3cm\epsfysize=3cm\epsfbox{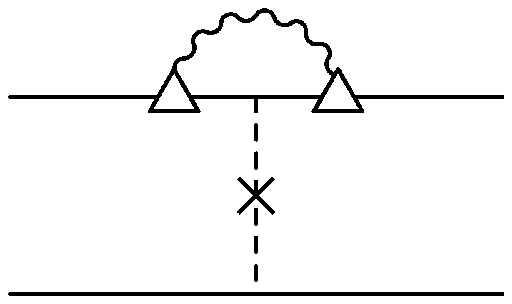}} (c)
\hspace{0.5cm}{\epsfxsize=3cm\epsfysize=3cm\epsfbox{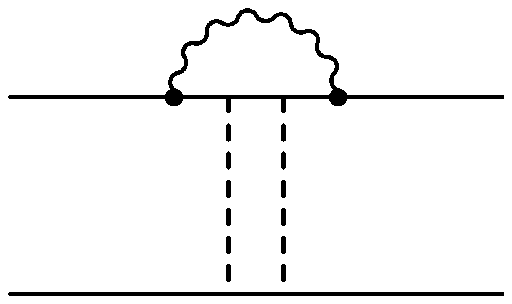}} (d)\\
\hspace{0.5cm}{\epsfxsize=3cm\epsfysize=3cm\epsfbox{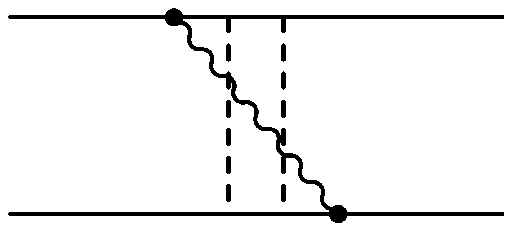}} (e)

\vspace{1cm}
\caption{Self-energy diagrams for the cooperon and diffuson.
The triangles denote impurity dressed vertices as in Fig.\ref{fig1}.
The dashed line with a cross is a single impurity line. Two dashed lines denote
an impurity ladder, i.e. a cooperon or diffuson.
Diagram (e) is not included in the calculation of the inelastic
scattering, but is included in the path integral approach to
the dephasing time.}
\label{fig4}
\end{figure}

In the previous section we calculated the inelastic scattering rate due to the
Coulomb interaction. 
We showed that the ``self-energy'' $\Sigma({\bf Q},\omega )$, which is given by
the sum of the 
diagrams shown in Fig.\ref{fig4},  is responsible of
various renormalization effects associated to the Coulomb interaction.
In particular, inelastic scattering arises from 
 only one type of these diagrams 
in the low energy region $(|\omega |< T)$ and within the
Matsubara formalism is  related to the 
existence of a branch cut contribution.
In the present section we address the determination of the dephasing time,
the time scale over which the quantum correction to the conductivity decays.
In order to do this calculation one needs to include also vertex corrections, 
i.e diagrams of the type shown in Fig.\ref{fig4}e.

The analysis is strongly simplified if we  restrict ourselves to
frequencies $|\omega| < T$ from the beginning. 
In this limit one neglects processes related to the diagrams
a,b,c and a part of d in Fig.\ref{fig4}.
In the case of the inelastic scattering time, as shown in the previous section,
this is immaterial since it only takes contributions from $| \omega |< T$.
Hence, by restricting the energy transfers to frequencies lower than the
temperature  some interaction effects are lost, but those  relevant for 
inelastic scattering and dephasing are taken into account. 

It is useful at this point to switch again to the real-time formalism. In the limit of 
low frequencies, the Keldysh component of the electron-electron interaction
given by $V^{K} = \coth( \omega /2 T ) ( V^R-V^A)$ becomes dominant due to
the $\coth ( \omega/2T )$ and the
retarded and advanced parts can be neglected. The Keldysh component of the
interaction has the same structure in Keldysh space as an external field, 
and thus the interaction
can be treated formally as a fluctuating external field.
At the end the fluctuation dissipation theorem is used to relate the field 
fluctuations to the internal electromagnetic field at thermal equilibrium.

With this in mind we start with the calculation of the dephasing rate for WL within 
a path-integral 
approach, following the literature\cite{aak82,eiler84}. 
The method allows us to include the vertex corrections in an elegant way.  
Then it will be straightforward to calculate the dephasing rate for EEI.

The particle-particle propagator obeys a diffusion equation whose
solution can be written in terms of a path integral
\begin{equation}
C^t_{\eta ,\eta '}({\bf r}, {\bf r}')={1\over 2 \tau }
\int^{{\bf r}_\eta ={\bf r}}_{{\bf r}'_{\eta '}={\bf r}'}{\cal D}{\bf r}_{t_1 }
\exp -( S_0+S_1)
\end{equation}
with
\begin{eqnarray}
S_0 &=& \int^{\eta}_{\eta '}\D t_1
{{\dot{\bf r}^2_{t_1}}\over{2D}} \\
S_1 &=& -\I e\int^{\eta}_{\eta '}\D t_1
\dot{\bf r}_{t_1}
\cdot{\bf A}_t({\bf r}_{t_1},t_1)
.\label{12}
\end{eqnarray}
The vector potential
${\bf A}_t({\bf r},t_1)={\bf A}({\bf r},t+t_1/2)+{\bf A}({\bf r},t-t_1/2)$
is defined as the sum of the vector potentials seen by the two particles
entering the propagator. To make contact with the diagrammatic approach,
the potential at times $t\pm t_1/2$ corresponds to the retarded (advanced)
electron lines in the diagram. The vector potential describes the effect of the
electromagnetic field generated by the other electrons. 
In the case of a fluctuating field with only quadratic correlations,
one can average over the field fluctuations, leading to
\begin{eqnarray}
&S_1 \to S_{\rm int} = &\nonumber \\
&{e^2\over 2} 
\int^{\eta}_{\eta '}{\rm d}t_1 \int^{\eta}_{\eta '}{\rm d}t_2
\dot{ r}^i_{t_1}\dot{ r}^j_{t_2}
\langle { A}_t^i({\bf r}_{t_1},t_1) { A}_t^j({\bf r}_{t_2},t_2)
\rangle 
. \label{13}
&\end{eqnarray}
It is useful to define the average over diffusive paths as
\begin{equation}
\langle [ \cdots ] \rangle_{d} = \int {\cal D} {\bf r}_t [\cdots ] \exp(-S_0 )/
\int {\cal D} {\bf r}_t \exp(-S_0 )
,\end{equation}
which allows us to write the cooperon in the presence of the fluctuation field
as the product of the unperturbed cooperon times a perturbation which has to be averaged
over diffusive paths:
\begin{equation}
C^t_{\eta ,\eta '}({\bf r}, {\bf r}')={1\over 2 \tau }
\int^{{\bf r}_\eta ={\bf r}}_{{\bf r}'_{\eta '}={\bf r}'}{\cal D}{\bf r}_{t_1 }
\exp (-S_0 )\langle \exp( -S_{\rm int}) \rangle_d
.\end{equation}
Following Ref.\cite{eiler84} we then make the approximation
\begin{equation}\label{eq51}
\langle \exp( -S_{\rm int}) \rangle_d \to \exp ( -\langle S_{\rm int} \rangle_d )
.\end{equation}
In the diagrammatic language this approximation is equivalent to selecting a 
certain subset of graphs. As realized in Ref.\cite{eiler84} one selects diagrams where
interaction lines do not intersect and thus correspond to graphs (d) and (e)
in Fig.\ref{fig4}.
In some cases $\langle S_{\rm int } \rangle_{d}$ is linear in $\eta-\eta'$
and then the dephasing time may be easily read off,
$\langle S_{\rm int } \rangle_{d} = (\eta-\eta')/2\tau_\phi$.
In general we determine the dephasing time from the condition
$\langle S_{\rm int } \rangle_{d} = 1 $ for $\eta=-\eta' =\tau_\phi $.
We may then write
\begin{eqnarray}
S_{\rm int}&=&
{e^2\over  2}
\int^{\eta}_{\eta '}{\rm d}t_1 \int^{\eta}_{\eta '}{\rm d}t_2
\dot{ r}^i_{t_1}\dot{ r}^j_{t_2}
\sum_{{\bf q},\omega}
\langle { A}^{i} {  A}^j\rangle_{ {\bf q},\omega}\nonumber\\
&&\times \exp \left( {\I{\bf q}\cdot({\bf r}_{t_1}-{\bf r}_{t_2}) }\right)
\nonumber\\
&&\times 2 \left[\cos (\omega {{t_1-t_2}\over 2})
 +\cos (\omega {{t_1+t_2}\over 2})\right]
\label{14}
,\end{eqnarray}
where
$\langle A^i A^j \rangle_{ {\bf q}, \omega } $ is the electromagnetic
field correlator in reciprocal space.
The cosine term in $t_1-t_2$ refers to
correlation of vector potentials seen by the same electron at different times,
while the term in $t_1+t_2$ represents the correlation of the vector potentials
of two electrons. In the diagrammatic language of the previous section these
latter terms have not been considered.
The electromagnetic field fluctuations may be decomposed in the
longitudinal and transverse parts. Here we are interested in the effect of
Coulomb interaction and we then consider the longitudinal part only.
Using the fluctuation dissipation theorem one finds the electric field 
fluctuations for small frequencies as
($\omega < T $)
\begin{equation}
e^2 \langle E^i E^j \rangle_{ {\bf q}, \omega }=
- q^i q^j {2T\over \omega } \Im V^R({\bf q}, \omega ) 
.\end{equation}
Inserting the retarded interaction from eq.(\ref{eq15}) we arrive at
\begin{equation}
\langle { A}^{i} { A}^j\rangle_{ {\bf q},\omega}=
{{q^i q^j}\over q^2}{{T}\over {e^2N_0D\omega^2}}
\label{15}
.\end{equation} 
Note that the fluctuation dissipation theorem has to be used in the classical limit,
$ \omega < T$. 
As it is clear from the discussion in the beginning of this
section it would be inconsistent to take the quantum limit
\begin{equation}
\langle A^i A^j \rangle_{{\bf q}, \omega } =
- { q^i q^j \over \omega^2 e^2 } \coth\left( {\omega \over 2 T} \right)
\Im V^R({\bf q}, \omega )
,\end{equation} 
since this would immediately
introduce contributions from $\omega > T$. For example in equation (\ref{6}) for the
inelastic scattering time 
this would amount to keep the $\coth(\omega/2T)$ while 
neglecting the $\tanh(\omega/2T)$.
Various claims about the saturation of the dephasing time appear
to be related to this mistake
\cite{golubev98,mohanty97}. 

Next we will integrate eq.(\ref{14}) by parts  by using the relations 
\begin{equation}
\dot{ r}^i_{t_1}\dot{ r}^j_{t_2}
q^i q^j
\exp \left(\I {\bf q}\cdot({\bf r}_{t_1} -{\bf r}_{t_2})\right)=
\partial_{t_1}\partial_{t_2}
\exp \left(\I{\bf q}\cdot({\bf r}_{t_1}-{\bf r}_{t_2}) \right)
\label{16} 
\end{equation}
\begin{eqnarray}
& \partial_{t_1}\partial_{t_2}
\left[\cos (\omega {{t_1+t_2}\over 2})
 +\cos (\omega {{t_1-t_2}\over 2})\right]& \nonumber\\
& =-{{\omega^2}\over 4}
 \left[\cos (\omega {{t_1+t_2}\over 2})
 -\cos (\omega {{t_1-t_2}\over 2})\right]
&\label{17}
.\end{eqnarray}
The integration over the frequency $\omega$
may be extended to infinity when considering 
the asymptotic behavior $\eta \gg 1/T$ (which involves times 
$t_1$ and $t_2$ which are much longer than
$1/T$) and one finds $\delta$-functions
\begin{equation}
{1\over 2}\int^{\infty}_{-\infty}{{{\rm d}\omega }\over {2\pi }}
\cos (\omega {{t_1\pm t_2}\over 2})=\delta (t_1\pm t_2 )
\label{18}
.\end{equation}
As a result of inserting eqs.(\ref{15}-\ref{18}) in eq. (\ref{14})
one obtains
\begin{equation}
 S_{\rm int } =
{{T}\over {2 N_0 D}}\sum_{\bf q}{1\over { q}^2}
\int^{\eta}_{\eta '}{\rm d}t
 \left(1- e^{\I {\bf q}\cdot({\bf r}_t-{\bf r}_{-t})}\right)
.  \label{19}
\end{equation}
Eq.(\ref{19}) is the first central result of this section. It shows that the small
$q$ singularity due to long range Coulomb interaction is cut off.  
There is therefore no need to introduce the refined form eq.(\ref{eq40}) for the 
Coulomb interaction.
This happens
because the oscillating term, originating from correlation between different
particles, cancels with the unity factor stemming from correlations on the
same particle. In the diagrammatic language this is the cancellation between
self-energy and vertex corrections in the small $q$ limit.
To proceed further, one has to average the oscillating factor over the
diffusive paths. However to obtain the leading behavior is enough to
proceed as follows. The oscillating factor is different from zero only
when $|{\bf q}\cdot({\bf r}(t)-{\bf r}(-t))| < 1$. Then one may
substitute the expression in brackets with unity with the condition
that $Dq^2|t|> 1$ or $Dq^2 > \tau^{-1}_{\phi}$. We note that the integral
over $q$ is divergent in the ultraviolet limit. This is because 
we took the integral over $\omega$ from minus
infinity to plus infinity. We may correct this error remembering that
for typical frequencies $Dq^2 \sim \omega < T$ and then use this 
condition as an ultraviolet cut-off in the $q$ integral of eq.(\ref{19}).
Then 
\begin{equation}
1=
{{T}\over {2 N_0 D}}\sum_{\tau^{-1}_{\phi}<Dq^2 < T}{1\over { q}^2}
\int^{\tau_\phi}_{-\tau_\phi}{\rm d}t~ 1=
\tau_\phi {T \over 4 \pi D N_0 } \ln{ T\tau_{\phi}}
 \label{20}
\end{equation}
which yields the AAK result in eq.(\ref{eq41}).

The dephasing time in the particle-hole channel can be obtained
by an analogous calculation.
The particle-hole propagator is given by
\begin{equation}
D_{t,t'}^{\eta}({\bf r}, {\bf r'})= \frac{1}{\tau}
\int^{{\bf r}_t={\bf r}}_{{\bf r}_{t'}={\bf r}'}{\cal D}{\bf r}_{t_1}
\E^{-(S_0+ S_1)}
,\end{equation}
with
\begin{eqnarray}
& S_0 = \int^{t}_{t'}\D t_1
{ {\dot{\bf r}^2_{t_1}}\over 4D } & \\
& S_1 = -\I e\int^{t}_{t'}\D t_1
\dot{\bf r}_{t_1}
\cdot \left( {\bf A}({\bf r}_{t_1},t_1+ \eta/2) -
       {\bf A}({\bf r}_{t_1},t_1 -\eta/2) \right)
&\end{eqnarray}
This differs from the expression for the cooperon by a factor two in $S_0$, and
what is most important, the external field couples differently in $S_1$.

Averaging over the fluctuating vector potential and going through
the same steps as before for the particle-particle propagator we
find
\begin{eqnarray}
S_{\rm int} &= &{ 1 \over 2}
\int_{0}^{\eta} \D t_1 \int_{0}^{\eta} \D t_2
\sum_{\bf q} { T \over D N_0 q^2 }
\exp\left[ \I {\bf q} \cdot ( {\bf r}_{t_1} - {\bf r}_{t_2} )\right]
\nonumber\\
&& \times \left( 2 \delta(t_1-t_2) -
         \delta(t_1-t_2-\eta ) - \delta(t_1-t_2 +\eta )
\right)
\label{eq63}
.\end{eqnarray}
We set the integration limits as needed in the calculation
of $\Delta {\rm j}_{\rm EEI}$, eq.(\ref{EEI}), and 
considered the limit $ \eta T \gg 1$.
Details of the explicit evaluation of the dephasing
time in the particle-hole channel are reported
in Appendix A.
The first delta function in (\ref{eq63}) corresponds to correlations of the field
for the same particle (or hole), whereas the two other delta functions
correspond to correlations between a particle and a hole. 
From the integration limits it follows that these do not contribute
to $S_{\rm int}$. The low momentum singularity is only cut-off if
we calculate the field fluctuations more accurately,
\begin{equation} 
{1\over Dq^2 } \to {\kappa \over  q} 
{D \kappa q \over \omega^2 + ( D \kappa q )^2 }
\label{new65}
,\end{equation}
according to the expression for the screened Coulomb interaction 
in our calculation
of $1/\tau_{\rm inel}$ in eq.(\ref{eq40}).
We find that the dephasing rate $1/\tau_\phi$ in the particle-hole channel
equals the inelastic rate $1/\tau_{\rm inel}$, and is thus definitely not
the same as the dephasing rate in the particle-particle channel.

Our result that the inelastic rate is equal to the dephasing rate in the particle-hole
channel is non-generic and is due to the dephasing mechanism considered here.
The crucial approximation involved is that the interaction is local in time.
For an interaction with retardation effects the inelastic scattering rate
and the dephasing rate will not be the same.
With this regard
we want to mention that the processes considered here are not
the only intrinsic processes leading to inelastic scattering and dephasing.
Castellani et al \cite{castellani86} calculated the dephasing rate
adding to the diagrams shown in Fig.\ref{fig4} the corresponding Hartree
diagrams. The latter correspond to exchange of spin-fluctuations.
In two dimensions, a contribution to $1/\tau_\phi$ which is linear in $T$ has been
found, which corrects the result given here by a numerical factor.
Beyond electron-electron scattering also electron-phonon scattering
is a further source of inelastic scattering and dephasing.
Often the latter is dominant at high temperature with 
$1/\tau_\phi  \propto T^3$.
In general any type of low lying excitation with $\omega < T$ that couples to the
conduction electrons causes inelastic scattering and dephasing. In the zero temperature
limit, however, the number of such excitations goes to zero and so does the 
dephasing rate.

\section{Summary and discussion}
We have studied non-linear effects and dephasing of the quantum corrections to
the conductivity.
We worked within a time domain representation. In this representation
the weak localization correction to the current at time
$t$ is related to processes where a particle and a hole start diffusing at the same time
$(t-\eta)$ around a closed loop but in opposite directions.
We succeeded in deriving a new expression for the interaction correction to the current
which allowed us to calculate the latter in an arbitrary time
dependent electromagnetic field.
In contrast to weak localization, the size of the interaction correction
is not related to diffusion around a closed loop.
In this case the correction to the conductivity can be written in terms of the
amplitude describing the propagation of  a particle and an hole with the
particle moving from ${\bf x}_i $ at $(t-2\eta)$ to ${\bf x}_f$ at
$(t-\eta)$, and the hole propagating along the same path from
${\bf x}_i$ at $(t-\eta)$ to ${\bf x}_f$ at $t$.

Due to these different processes contributing to WL and EEI we find
substantially different non-linear effects.
In a static electric field, no non-linear effects were found for WL
whereas in case of EEI non-linear effects arise, 
when the voltage drop over a thermal length is of the order of $k_B T$.
Note that we did not take into account heating of the sample.
Experimental evidence for non-linear effects on EEI which are not related to 
heating has been found several years ago by Bergmann et al\cite{bergmann90} 
in gold films. Qualitative arguments why this effects should exist were 
given in that work, but a 
quantitative theory was not worked out.

Recently non-linear effects have been observed 
in two dimensional systems\cite{kravchenko96,yoon99} 
near the metal-insulator transition.
The physical origin of the metal-insulator transition and the metallic phase
in these systems is not clear, 
but one candidate is quantum interference in the interaction triplet
channel\cite{castellani98}. In this case
one would expect to observe in the metallic phase the non-linear effects 
which we have described in this paper.  
The sign of the effect would depend on the relative importance of the singlet
and triplet contributions to the EEI corrections\cite{note}. 
However, from the available experimental data 
deep in the metallic phase
we are not able to separate non-linear effects due to heating from those due
 to interference,
so at present we cannot provide a quantitative comparison of theory and 
experiment. 

Besides the effect of a static electric field we examined the quantum
corrections in the presence of a microwave field.
We confirmed earlier calculations on the effect of a microwave field  on WL and
worked out the effects on EEI. 
A microwave field of arbitrarily low frequency can destroy
EEI efficiently, if the voltage drop on a thermal length is of order of the
temperature, similar to the non-linear effects in a static field.
A high frequency microwave field ($\hbar \omega > k_B T$)  
is less effective in destroying
EEI, i.e. EEI increases when increasing the microwave frequency. 
This behavior has been observed experimentally 
by Liu and Giordano\cite{liu91} in highly disordered gold films. 
In those experiments the change of the DC conductivity with 
the amplitude of the microwave field
was measured, $\sigma(E_{\rm AC} ) \propto A E_{\rm AC}^2$, and the factor $A$ was
found to be frequency dependent.
In Ref.\cite{liu91} a frequency dependence of the interaction correction
to the conductivity was suggested as a possible
origin of the experimental results. 
Whereas experimental results are qualitatively in agreement with our theory 
(see Fig.\ref{fig3c}) the experimentally observed effect is much larger 
(by a factor of about $10^3$) than predicted theoretically.

Vitkalov et al\cite{vitkalov88} studied the conductivity of 
Si-MOS transistors in the presence of a microwave field. 
For electron densities above $n_s > 10^{12}/{\rm cm}^2$ 
the experimental results agreed with 
the theoretical considerations on weak localization in the presence 
of a microwave field. At lower 
density, $2 \cdot 10^{11}/{\rm cm}^2 <n_s < 10^{12}/{\rm cm}^2$, 
they found non-heating effects
which were incompatible with weak localization. 
From the parameters given in\cite{vitkalov88} 
we conclude that non-linear effects related to 
electron-electron interaction become important 
at low density as we demonstrate here below. 
In a weak microwave field of low frequency $\omega$, 
the weak localization correction  
is quadratic in the electric field and given by\cite{altshuler85} 
$\sigma_{\rm WL}(E_{\rm AC} ) -\sigma_{\rm WL}(0 )\approx 
0.13 \cdot (e^2/\pi h) D(e E_{\rm AC})^2 \omega^2\tau_\phi^5$. 
The interaction correction on the other hand is
$\sigma_{\rm EEI}(E_{\rm AC}) -\sigma_{\rm EEI}(0) \approx 0.08 \cdot (e^2/\pi h)  
 \hbar^3 D(e E_{\rm AC})^2/(k_B T)^3 $.
Inserting the parameters given in Ref.\cite{vitkalov88} at density 
$n_s=2 \cdot 10^{12}/{\rm cm}^2$, $T=4.2$K, 
$\tau_\phi =0.5 \cdot 10^{-11}$s which corresponds to $\hbar/\tau_\phi \approx 1.5$K,
and $\hbar \omega \approx  0.5$K, it is seen 
that WL and EEI are comparable
in size. 
Note the strong dependence of $ \sigma_{\rm WL}$ on the dephasing time, so varying
$\tau_\phi$ by two varies the $\sigma_{\rm WL}$ by almost two orders of magnitude.
EEI dominates if the dephasing time becomes shorter, as  is the case for
lower electron density. In that regime non-linear effects are only
weakly magnetic field dependent (due to Zeeman splitting in the spin-triplet channel which 
we did not consider in this paper),  and
proportional to $1/T^3$, 
so by reducing the temperature from 4.2K to 1.7K the non-linearities are expected
to increase by more than one order of magnitude.  
Both statements agree with the experimental observations.

Finally, we considered dephasing due to intrinsic electric
field fluctuations. For WL we confirmed the AAK result, leading
to $1/\tau_\phi  \propto T$ in two dimensions. In case of EEI
we found, as it was suggested earlier by Castellani et al\cite{castellani86}, that
there is dephasing also in the particle-hole channel. However in contrast
to this earlier suggestion we found different dephasing rates in the 
particle-hole and particle-particle channels, both of which however vanish as $T \to 0$.
Mohanty et al \cite{mohanty97} observed recently a 
saturation of the weak localization dephasing time in gold wires. 
It was shown in the experiments
that the saturation was not due to heating or magnetic impurities.
The dephasing rate saturated below 
$T \sim 1$K to values of the order $\hbar/\tau_\phi \sim 1$--10mK.
We found that the interaction correction is affected by dephasing, so when
weak localization saturates, also the interaction correction is expected to saturate at 
still lower temperature.
In the samples with the highest saturation values of $1/\tau_{\phi}$  
a saturation of the interaction correction to the conductivity 
was indeed observed\cite{mohanty97} below $T\sim 100$mK. 
There remains the task to determine the mechanism causing saturation. Clearly it cannot be
equilibrium electric field fluctuations.     
Also experimental evidences are against the idea of such an intrinsic
decoherence mechanism\cite{gershenson98}.
Altshuler et al\cite{altshuler98} suggested an external (non-equilibrium) 
microwave field as the
origin of the saturation of the weak localization dephasing time.
The weak localization dephasing time, $\tau_\phi= \tau_{\rm AC}$, 
and the saturation temperature of EEI, $T_{\rm sat}$, 
are functions of the amplitude
and frequency of the microwave field.
It was pointed out by Altshuler et al that 
a microwave field  is most efficient in destroying weak localization, when
the frequency is close its ``optimum'' value i.e., 
the inverse dephasing time, $\omega \sim 1/\tau_{\rm AC}$, which
leads to 
$1/\tau_{\rm AC} \sim [D (e E_{\rm AC}/\hbar)^2]^{1/3}$. 
In this case we expect saturation of
EEI at $ k_B T_{\rm sat} \sim  \hbar/\tau_{\rm AC}$. 
The experimental saturation temperature
however is larger and roughly 
found as $k_B T_{\rm sat } \sim 10 \hbar/\tau_{\rm AC}$.  
This could be explained with a microwave frequency which is not ``optimum''. 
By decreasing the frequency, the WL dephasing time becomes larger
according to $1/\tau_{\rm AC} \approx [D(e E_{\rm AC}/\hbar)^2 \omega^2]^{1/5}$,
whereas the saturation temperature of EEI does not depend on frequency.  
From $k_B T_{\rm sat} \sim 10 \hbar/\tau_{\rm AC}$ 
we estimate the frequency which is consistent
with the experimentally observed saturation of WL and EEI as 
$\hbar \omega \sim k_B T_{\rm sat}/10^{5/2}$.

On the other hand saturation of the resistance could also occur due to heating.
In Ref.\cite{altshuler98} strong heating 
is assumed to set in when the voltage drop over the
length of the wire is of order of the temperature.
From our considerations we found a saturation of the resistance
when the voltage drop over a thermal length
is of the order of the temperature.
In the experiments the thermal length was much shorter
than the system size, so the saturation of the resistance 
is most probably due to heating if the conditions of Ref.\cite{altshuler98}
are satisfied. 
This however should be checked experimentally.

The observed saturation of dephasing an resistance 
could also be caused by   
some intrinsic processes, other than
the equilibrium electric field fluctuations.
Recently it has been suggested that two-level-systems may lead to a temperature independent
weak localization dephasing rate\cite{imry99,zawadowski99}. 
To assess the agreement with the experiment one has to calculate the
dephasing rate in the particle-hole channel using for example the formalism 
we developed in this article. 
This however is beyond the scope of the present paper.
\acknowledgements
This work was supported by MURST under contract no. 9702265437 (R.R.) and
the TMR program by the European Union (P.S.).  
C.C. and R.R. acknowledge useful conversation with M. Sarachik and
S. Vitkalov. 
We thank M.Leadbeater for his help in the preparation of the final
version of the paper.
\begin{appendix}
\section{Dephasing in the particle-hole channel}
In this appendix we report the detailed calculation 
of the correction to the particle-hole
propagator caused by
electric field fluctuations.  
We start from the path integral representation of the diffuson, after
averaging over the fluctuating field and after the approximation (\ref{eq51}): 
\begin{equation}
D^\eta_{t-\eta/2, t-3\eta/2 }({\bf r}, {\bf r})
= {1\over \tau} \int {\cal D} {\bf r}_t 
\exp(- S_0-\langle S_{\rm int} \rangle_d ) 
,\end{equation}
where $S_0$ has been defined in section V.
The relevant term for dephasing is $S_{\rm int}$ and is given by
\begin{eqnarray}\label{eqA2}
S_{\rm int} &=& \int_0^\eta \D t_1 \int_0^\eta \D t_2
               \int_{-T}^T {\D\omega \over 2\pi} \sum_{\bf q}
              {T \over DN_0 } {D^2 \kappa^2 \over \omega^2 + (D \kappa q )^2}\nonumber\\
            && \times   \exp[ \I {\bf q}( {\bf r}_{t_1} -{\bf r}_{t_2}) 
                  -\I\omega (t_1 -t_2) ]
           \left[ 1 -\cos(\omega\eta) \right]
.\end{eqnarray}    
The $\cos(\omega\eta)$ term is due to the vertex corrections.
The expression we gave in eq.(\ref{eq63}) is recovered when
neglecting the frequency dependence of the field fluctuations
 and considering large times
$\eta$ only, such that the $\omega$-integration leads to delta
 functions in time.  
Here we do not rely on these approximations.
First we average (\ref{eqA2}) over diffusive paths,
$\langle \exp[ \I {\bf q}( {\bf r}_{t_1} -{\bf r}_{t_2}) ]\rangle_d  \approx
\exp( - Dq^2|t_1-t_2 | ) $ and switch to ``center of mass''  
and ``relative'' variables 
in the time integrations,
$ t_+ = (t_1 +t_2)/2$ and $t_- = t_1 -t_2$, leading to
\begin{eqnarray} 
\langle S_{\rm int} \rangle_d&=&  2 \int_0^{\eta/2} \D t_+ \int_{-2t_+}^{2t+} \D t_-
               \int_{-T}^T {\D\omega \over 2\pi} \sum_{\bf q}
              {T \over DN_0 } {D^2 \kappa^2 \over \omega^2 + (D \kappa q )^2}\nonumber\\
            && \times   \exp( Dq^2 |t_-| -\I\omega t_- )
           \left[ 1 -\cos(\omega\eta) \right]
.\end{eqnarray}
Note that the time integration in this expression covers only half of the
region of the original integral. We cured this by multiplication by two.
Then it is straightforward to integrate over $t_-$ and $t_+$,
$$
\int_0^{\eta/2} \D t_+ \int_{-2t_+}^{2t_+} \D t_- \exp( - Dq^2|t_- | -\I \omega t_- )
$$
\begin{equation}\label{eqA4}
= \Re \left\{
{\eta \over \I \omega + Dq^2 } + \left( {1\over \I \omega +Dq^2} \right)^2
\left[\E^{-(Dq^2+\I \omega ) \eta } -1 \right] \right\}
.\end{equation} 
For large momenta ($Dq^2 >1/\eta $) the second term is negligible, so
(\ref{eqA4}) is proportional to the inverse diffusion propagator.
The second term becomes important in the low momentum region, and cancels the
low-$q$ low-$\omega$ singularity of the propagator. Explicitely we find 
\begin{equation}
({\rm \ref{eqA4}}) = \left\{ \begin{array}{ll}
\eta Dq^2/[ \omega^2 + (Dq^2 )^2 ]            & {\rm for }\,\, Dq^2 >1/\eta \\
\eta [Dq^2+(1-\cos\omega\eta)/\eta]/\omega^2 & {\rm for }\,\, Dq^2< 1/\eta, \omega \eta > 1  \\
\eta^2/2                                      & {\rm for }\,\, Dq^2 <1/\eta, \omega \eta < 1
\end{array}
\right.
.\end{equation}
Calculating $\langle S_{\rm int} \rangle_d$ for large times $\eta \gg 1/T$ 
we find
\begin{equation}
\label{a6}
\langle S_{\rm int} \rangle_d =  {T\eta \over 4\pi D N_0 }
 \ln ( D\kappa^2 T \eta^2) 
,\end{equation} 
from the region of the momentum integration with
 $Dq^2 < T$ and $D\kappa q > 1/\eta$.
Setting $\langle S_{\rm int} \rangle_d =1$ at $\eta=\tau_\phi$
reproduces the FA result for the
inelastic scattering time, see eq.(\ref{eq43}). 
In the relevant region of the momentum integration
($D\kappa q > 1/\eta $) the electric field fluctuations 
are nearly local in time. 
As a consequence
vertex corrections are not important in this region. 

The opposite limit, $\eta \ll 1/T$, which is relevant
for $\tau_\phi \ll 1/T$,
may be more interesting since only small times 
$\eta$ contribute to the interaction correction to the current.
In this limit the vertex corrections are important, 
and the dephasing time is strongly reduced below
the inelastic time, since
$1-\cos(\omega\eta) \approx (\omega\eta)^2/2$, compare (\ref{eqA2}). 
However also in this region it is essential to take the refined form (\ref{eq40})
of the Coulomb interaction in order to determine the electric field fluctuations.
The vertex corrections do not cancel the low-$q$ singularity of $\Im V(q)$.
Going explicitly through the algebra again, we find
\begin{equation} \label{eqA7}
\langle S_{\rm int} \rangle_d =
 {  T^4 \eta^4 \over 24 \pi^2 D N_0} \ln {D\kappa^2 \over T^2 \eta }
.\end{equation} 
The relevant region in momentum space is $Dq^2 < 1/\eta$, $D\kappa q > T$.
From (\ref{eqA7}) the dephasing time in the limit where 
$\tau_\phi \ll 1/T$ is determined as
\begin{equation}
{1\over \tau_\phi } = T \left( { 1\over  24\pi^2 D N_0 }
 \ln{ D\kappa^2 \over T^2 \tau_\phi }
\right)^{1/4}=
T \left( { 1\over  24\pi^2 D N_0 }
 \ln{ {D\kappa^2 \over T }(24\pi^2N_0D)^{1/4} }
\right)^{1/4}.\end{equation}

\end{appendix}



\begin{references}
\bibitem{altshuler85}
                B.L. Altshuler, A.G. Aronov, D.E. Khmelniskii, and A.G. Larkin,
                in {\em Quantum Theory of Solids}, edited by I.M. Lifshitz
                (MIR Publishers, Moscow, 1982);
                B.L. Altshuler and A.G. Aronov, 
                in {\em Electron-Electron Interactions in Disordered Systems},
                edited by M. Pollak and A.L. Efros (North-Holland, 
                Amsterdam, 1985), p. 1.
\bibitem{lee85} P.A. Lee and T.V. Ramakrishnan,
                Rev. Mod. Phys. {\bf 57}, 287 (1985);
		D. Belitz and T.R. Kirkpatrick,
		Rev. Mod. Phys. {\bf 66}, 261 (1994).
\bibitem{golubev98}D.S. Golubev and A.D. Zaikin,
		Phys. Rev. Lett. {\bf 81}, 1074 (1998).
\bibitem{critic}D. Cohen, and Y. Imry,  preprint, cond-mat/9807038;
                K.A. Erikson and P. Hedegard, preprint, cond-mat/9810297;
                M. Vavilov and V. Ambegaokar, preprint, cond-mat/9902127.
\bibitem{altshuler98}B.L. Altshuler, M.E. Gershenson, and A.L. Aleiner,
                Physica E {\bf 3}, 58 (1998); A.L. Aleiner, B.L. Altshuler, and
		M.E. Gershenson, preprint, cond-mat/9808053.
\bibitem{schmid74} A. Schmid, Z. Phys. {\bf 271}, 251 (1974).
\bibitem{abrahams81}E. Abrahams, P.W. Anderson, P.A. Lee, T.V. Ramakrishnan,
                Phys. Rev. B {\bf 24}, 6783 (1981).
\bibitem{aak82} B.L. Altshuler, A.G. Aronov, and D.E. Khmelnitsky,
                J. Phys. C:Solid State Phys. {\bf 15}, 7367 (1982).
\bibitem{fa83}  H. Fukuyama and E. Abrahams, Phys. Rev. B {\bf 27}, 5976 (1983).
\bibitem{fukuyama84}
                H. Fukuyama,
                J. Phys. Soc. Jpn. {\bf 53}, 3299 (1984).
\bibitem{eiler84}W. Eiler,
                J. Low Temp. Phys. {\bf 56}, 481 (1984).
\bibitem{aronov84}A.G. Aronov, Physica B {\bf 126}, 314 (1984).
\bibitem{abrahams85}E. Abrahams in {\em Localization and Metal-Insulator
                      Transition}, edited by H. Fritsche and D. Adler (Plenum,
                     New York, 1985), p.433.
\bibitem{castellani86}C. Castellani, C. Di Castro, G. Kotliar, and P.A. Lee,
                       Phys. Rev. Lett. {\bf 56}, 1179 (1986).
\bibitem{stern90}A. Stern, Y. Aharonov, and Y. Imry,
                Phys. Rev. A {\bf 41}, 3436 (1990).
\bibitem{blanter96}U. Sivan, Y. Imry, and A.G. Aronov,
                Europhys. Lett. {\bf 28}, 115 (1994);
                Ya.M. Blanter, Phys. Rev. B {\bf 54}, 12807 (1996);
		B.L. Altshuler, Y. Gefen, A. Kamenev, and L.S. Levitov,
		Phys. Rev. Lett. {\bf 78}, 2803 (1997).
\bibitem{lin87}J.J. Lin and N. Giordano, Phys. Rev. B {\bf 35}, 1027 (1987).
\bibitem{pooke89}D.N. Pooke, N. Paquin, M. Pepper, and A. Gundlach,
                J. Phys. Condens. Matter {\bf 1}, 3289 (1989).
\bibitem{hiramoto89}T. Hiramoto et al, Appl. Phys. Lett. {\bf 54}, 2103 (1989).
\bibitem{mueller94}R.M. Mueller, R. Stasch, and G. Bergmann,
    		Solid State Commun. {\bf 91}, 255 (1994).
\bibitem{mohanty97} P. Mohanty, E.M.Q. Jariwala, and R.A. Webb,
                Phys. Rev. Lett. {\bf 78}, 3366 (1997);
                P. Mohanty, and R.A. Webb, Phys. Rev. B{\bf 55}, 13452 (1997);
                P. Mohanty, E.M.Q. Jariwala, and R.A. Webb,
                Fortschr. Phys. {\bf 46}, 779 (1998).
\bibitem{ward}  We notice that the inclusion of the vertex diagrams is also
                required by Ward identities, which implement the particle
		conservation and gauge invariance. Within the
		quasiclassical point of view inclusion of the vertex diagrams is
		equivalent to the statement that the phase {\em difference}
                of two paths determines dephasing.
\bibitem{bergmann90}G. Bergmann, Wei Wei, Yao Zou, and R.M. Mueller,
                Phys. Rev. B {\bf 41}, 7386 (1990).
\bibitem{kravchenko96}S.V. Kravchenko, D. Simonian, M.P. Sarachik,
                W. Meson, and G.E. Fourneaux,
		Phys. Rev. Lett. {\bf 77}, 4938 (1996).
\bibitem{yoon99}J. Yoon, C.C. Li, D. Shahar, D.C. Tsui, and M. Shayegan,
		Phys. Rev. Lett. {\bf 82}, 1744 (1999).
\bibitem{liu91} J. Liu and N. Giordano, Phys. Rev. B {\bf 43}, 1385 (1991).
\bibitem{vitkalov88}S.A. Vitkalov, G.M. Gusev, Z.D. Kvon, G.I. Leviev, and V.I. Fal'ko,
                Sov. Phys. JETP {\bf 67}, 1080 (1988).
\bibitem{rammer86}J. Rammer and H. Smith,
                Rev. Mod. Phys. {\bf 58}, 323 (1986).
\bibitem{altshuler80}B.L. Altshuler, A.G. Aronov, and P.A. Lee,
                Phys. Rev. Lett. {\bf 44}, 1288 (1980);
		B.L. Altshuler, D. Khmel'nitzkii, A.I. Larkin, and P.A. Lee,
		Phys. Rev. B {\bf 22}, 5141 (1980).
\bibitem{castellani84}C. Castellani, C. Di Castro, P.A. Lee, and
                M. Ma, Phys. Rev. B {\bf 30}, 527 (1984).
\bibitem{castellani98}C. Castellani, C. Di Castro, and P.A. Lee,
		Phys. Rev. B {\bf 57}, R9381 (1998).
\bibitem{note}  In this paper we have not considered the quantum interferences 
                in the interaction
                triplet channel. However we can obtain analogous results 
                in the triplet channel as those discussed 
                in this paper for the singlet channel.
\bibitem{gershenson98}Yu.B. Khavin, M.E. Gershenson, and A.L. Bogdanov,
                Phys. Rev. Lett. {\bf 81}, 1066 (1998).
\bibitem{imry99}Y. Imry, H. Fukuyama and P. Schwab, preprint, cond-mat/9903017.
\bibitem{zawadowski99}A. Zawadowski, J. von Delft, and D.C. Ralph, preprint,
		cond-mat/9902176.
\end{references}
\end{document}